\documentclass[desactivate]{aa}
\usepackage[varg]{txfonts}
\usepackage{graphicx}
\graphicspath{{images/}} 
\usepackage[version=3]{mhchem} 
\usepackage{pifont} 
\usepackage{longtable}
\bibpunct{(}{)}{;}{a}{}{,} 
\usepackage{color}
\AddToHook{begindocument/before}{\RequirePackage{hyperref}}

\begin{document}

	\title{The factors that influence protostellar multiplicity} 
	\subtitle{I. Gas temperature, density, and mass in Perseus with Nobeyama}
	
	\author{N. M. Murillo\inst{1,2}, C. M. Fuchs\inst{3}, D. Harsono\inst{4}, N. Sakai\inst{1}, A. Hacar\inst{5}, D. Johnstone\inst{6,7}, R. Mignon-Risse\inst{8,9}, S. Zeng\inst{1}, T-.H. Hsieh\inst{10}, Y-.L. Yang\inst{1}, J. J. Tobin\inst{11} \and M. V. Persson\inst{12}}
	
	\institute{Star and Planet Formation Laboratory, RIKEN Cluster for Pioneering Research, Wako, Saitama 351-0198, Japan\\ \email{nmurillo@astro.unam.mx}
		\and Instituto de Astronomía, Universidad Nacional Autónoma de México, AP106, Ensenada CP 22830, B. C., México
		\and Fox2Space - FTSCO, The Fault Tolerant Satellite Computer Organization, Weigunystrasse 4, 4040 Linz, Austria
		\and Institute of Astronomy, Department of Physics, National Tsing Hua University, Hsinchu, Taiwan
		\and Department of Astrophysics, University of Vienna, Türkenschanzstrasse 17, A-1180 Vienna
		\and NRC Herzberg Astronomy and Astrophysics, 5071 West Saanich Rd, Victoria, BC, V9E 2E7, Canada
		\and Department of Physics and Astronomy, University of Victoria, Victoria, BC, V8P 5C2, Canada
		\and Institutt for Fysikk, Norwegian University of Science and Technology, Høgskloreringen 5, Trondheim, 7491, Norway
		\and Universit\'{e} Paris Cit\'{e} CNRS, CNES, Astroparticule et Cosmologie, F-75013 Paris, France
		\and Max-Planck-Institut f\"{u}r extraterrestrische Physik, Giessenbachstrasse 1, D-85748 Garching, Germany
		\and National Radio Astronomy Observatory, Charlottesville, VA 22903, USA
		\and Independent researcher, Sweden}
	
	\abstract
	{Protostellar multiplicity is common at all stages and mass ranges. However, the factors that determine the multiplicity of protostellar systems have not been systematically characterized through their molecular gas.}
	{This work seeks to characterize the physical properties of the Perseus Molecular Cloud at $\geq$5000 AU scales through mapping of diagnostic molecular lines.}
	{Nobeyama 45m Radio Observatory (NRO) on-the-fly maps of \ce{HCN}, \ce{HNC}, \ce{HCO+}, and \ce{N2H+} ($J$ = 1--0) toward five subregions in Perseus, complemented with single pointing Atacama Pathfinder EXperiment (APEX) observations of \ce{HNC} ($J$ = 4--3) are used to derive physical parameters of the dense gas. Both observations have spatial resolutions of $\sim$18$\arcsec$, equivalent to $\sim$5000 AU scales at the distance of Perseus. Kinetic gas temperature is derived from the $I$(\ce{HCN})/$I$(\ce{HNC}) $J$ ratio, and \ce{H2} density is obtained from the \ce{HNC} $J$=4--3/$J$=1--0 ratio. These parameters are used to obtain the \ce{N2H+} (cold) and \ce{HCO+} (warm) gas masses. The inferred and derived parameters are then compared to source parameters, including protostellar multiplicity, bolometric luminosity and dust envelope mass.}
	{Inferred mean kinetic gas temperature ($I$(\ce{HCN})/$I$(\ce{HNC}) $J$=1--0 ratio; ranging between 15 and 26 K), and \ce{H2} volumetric density (\ce{HNC} $J$=4--3/$J$=1--0; 10$^{5}$ -- 10$^{6}$ cm$^{-3}$) do not show correlations with multiplicity in Perseus. The derived gas and dust masses, 1.3 to 16 $\times~10^{-9}$ M$_{\odot}$ for the cold gas mass (\ce{N2H+}), 0.1 to 25 M$_{\odot}$ for envelope dust masses (850 $\mu$m), and  0.8 to 10 $\times~10^{-10}$ M$_{\odot}$ for the warm gas mass (\ce{HCO+}), are correlated to multiplicity and number of protostellar components. The warm gas masses are a factor of 16 lower than the cold gas masses.}
	{This work shows that gas and dust mass is correlated to multiplicity at $\sim$5000 AU scales in Perseus. Higher order multiples tend to have higher gas and dust masses in general, while close binaries (separations $\leq$7$\arcsec$) and single protostars have similar gas and dust mass distributions. On the other hand, \ce{H2} density and kinetic gas temperature do not show any correlation with multiplicity.}
	
	
	\titlerunning{Perseus gas temperature and multiplicity}
	\authorrunning{Murillo et al.}
	
	\maketitle
	
	\section{Introduction}
	\label{sec:intro}
	Several high spatial resolution surveys of dust continuum have characterized the statistics of multiple protostellar systems \citep{looney2000,chen2013,tobin2016,encalada2021,tobin2022}.
	These surveys provide information on multiplicity fraction, companion fraction, distribution of components within systems, and speculate on formation scenarios based only on dust continuum.
	One of the missing pieces of information are the physical conditions of the gas surrounding the multiple systems. 
	In this sense, large-scale mapping of molecular gas is necessary to understand the processes and physical conditions that are occurring within the protostellar cloud cores \citep{mairs2014,mairs2016}.
	
	A pilot sample of 10 protostellar systems in Perseus was observed with the Atacama Pathfinder EXperiment (APEX) with to study the temperature-multiplicity relation at envelope scales ($\sim$7000 AU; \citealt{murillo2018}).
	The results of the pilot study showed that temperature was not correlated with multiplicity, while cold gas reservoirs might be more relevant to multiplicity.
	But the small sample size did not allow a systematic view.
	Tracing the physical conditions of the gas from the molecular cloud down to the protostellar disk for a large sample of systems is necessary in order to determine whether the previously established correlations apply globally for the Perseus molecular cloud or not.
	
	Molecular line observations allow the physical conditions to be probed.
	Gas kinetic temperature at molecular cloud scales can be indirectly derived from the $I$(\ce{HCN})/$I$(\ce{HNC}) $J$=1--0 ratio (e.g., \citealt{graninger2014,hacar2020,pazukhin2022}), found to be effective in the 15 to 40 K range from observations \citep{hacar2020}. 
	Another often used tracer is ammonia \ce{NH3} (e.g., \citealt{friesen2017,keown2019}), previously used to calibrate the $I$(\ce{HCN})/$I$(\ce{HNC}) ratio \citep{hacar2020,pazukhin2022}.
	Dust temperature maps have been derived through modified black-body fitting of spectral energy distributions (SEDs; e.g., \citealt{lombardi2014,zari2016}).
	This enables a comparison between gas and dust temperatures which is important as both are not necessarily thermalized.
	Gas density and mass  can be probed by \ce{N2H+}, a cold and dense gas tracer, and \ce{HCO+} which typically traces warm dense gas (e.g., \citealt{hsieh2019,murillo2022b}). 
	Dust envelope mass can be estimated from the 850 $\mu$m peak continuum emission (e.g., \citealt{jorgensen2004,murillo2016}).
	Combining these physical parameters with information on protostellar multiplicity from continuum surveys (e.g., \citealt{tobin2016}), source parameters derived from SEDs (e.g., \citealt{murillo2016}), and outflow directions (e.g., \citealt{stephens2017}) the factors that influence protostellar multiplicity from molecular cloud scales can be studied.
	
	A correlation between the \ce{N2H+} and \ce{NH3} column densities was found in Perseus starless and protostellar cores, but not between \ce{N2H+} and core density \citep{johnstone2010}. 
	A similar result was obtained for the entire NGC1333 region \citep{hacar2017}.
	Analysis of dust continuum masses versus thermal and non-thermal Jeans mass at different scales, and considering the multiplicity of protostars, suggest that thermal support is significant at small scales (few 1000 AU) but not at molecular cloud scales ($\geq$10 pc; \citealt{pokhrel2018}), however, the study assumed a single temperature for all cores.
	Massive cloud cores ($\geq$40 M$_{\rm \odot}$) containing fragments with average masses in the range of 2 to 4 M$_{\rm \odot}$ hint at the role of magnetic fields in controlling the mass and size of fragments within cloud cores, as well as the distribution of fragments \citep{palau2021}.
	Comparison of protostellar multiplicity in Perseus to dust core morphology and binary formation scenarios suggests that wide binaries either inspiral to become tight binaries or one of the components is ejected \citep{sadavoy2017}.
	Dust polarization and molecular gas observations of Class 0 systems found magnetic fields aligned with outflow axis and low angular momentum in the inner envelope for single sources \citep{galametz2020}.
	
	The multiplicity resulting from fragmentation is affected by mass, temperature, feedback or turbulence as discussed in recent reviews \citep{krumholz2014,lee2020,offner2022ppvii}.
	Simple analytic models suggest that structure dictates fragmentation, and a cloud core with a few Jeans masses can already fragment \citep{kratter2008,pon2011,pon2012}.
	Early models suggested that radiative feedback and gas heating could suppress core fragmentation by increasing the thermal Jeans mass (30 K at a few thousand AU: \citealt{krumholz2006}; extensive envelope heating out to a thousand AU scales: \citealt{bate2012}), even during the first collapse of the core \citep{boss2000,whitehouse2006}.
	Other models suggest that radiative heating mainly influences disk scales (few 100 AU) rather than the cloud core (\citealt{offner2022ppvii} and references therein).
	Self-consistent hydrodynamical simulations that include mechanical feedback through outflows (e.g., \citealt{guszejnov2021,mathew2021}) seem to show that outflows influence the fragmentation process and the physical conditions of the cloud, consistent with observations (e.g., \citealt{vankempen2009,yildiz2015}).
	The combination of magnetic fields and radiation also show an impact, to a degree, in the core fragmentation process \citep{commercon2010,hennebelle2011}, while the ratio of the rotational to magnetic energy of the core is also found to control core fragmentation in the innermost regions (\citealt{machida2008} in the low-mass case, \citealt{mignon-risse2023} in the high-mass case).
	Some models suggest that magnetic fields may enhance multiplicity (e.g., \citealt{offner2016,lee2019,mathew2021}) but reduce higher-order multiple protostellar systems.
	On the other hand, non-ideal magnetic processes are found to not inhibit protostellar multiplcity (e.g., \citealt{wurster2019}).
	Turbulent fragmentation is another proposed mechanism for the formation of multiple protostellar systems (e.g., \citealt{offner2010,padoan2014,federrath2015,cunningham2018,lee2019}).
	Models of high-mass star formation find that turbulence, which carries angular momentum, promotes disk growth and subsequent fragmentation, whereas magnetic fields reduce initial fragmentation \citep{commercon2011} and remove angular momentum on disk-scales (via magnetic braking and outflows), thereby reducing disk fragmentation \citep{mignon-risse2021}.

	To give observational insight into the molecular gas environment around multiple protostellar systems, this work presents single-dish observations toward five subregions in the Perseus Molecular cloud.
	The observations include on-the-fly (OTF) maps and single-pointing observations toward a sample of 37 protostellar systems.
	Given the spatial coverage of the OTF maps, a few starless cores are also included in the sample which serve to further reveal the physical conditions of cores that form multiple protostellar systems.
	A brief overview of the Perseus molecular cloud and sample selection criteria are given in Section~\ref{sec:sample}, with source tables given in Appendix~\ref{ap:obsdata}. 
	Observations are described in Section~\ref{sec:observations}, along with details on frequency setup, data reduction, and imaging. 
	Additional details are given in Appendices~\ref{ap:obsdata} and \ref{ap:moment}.
	Section~\ref{sec:results} briefly describes the resulting maps and spectra, with more detailed descriptions provided in Appendix~\ref{ap:moment}.
	Derivation of physical parameters, statistics and discussion on physical parameters versus multiplicity are described in sections~\ref{sec:analysis} and \ref{sec:discussion}.
	Conclusions are given in Section~\ref{sec:conclusions}.
	
	\section{Sample}
	\label{sec:sample}
		
	The Perseus molecular cloud presents a wide range of star forming environments with distances varying between 279 to 302 pc \citep{zucker2018}.
	The molecular cloud has been studied in molecular line emission (e.g. \citealt{arce2010,curtis2010a,curtis2010b,curtis2011,walker2014,hacar2017,tafalla2021,dame2023}), dust continuum (e.g. \citealt{hatchell2005,enoch2006,chen2016,pokhrel2018}), and dust polarization (e.g. NGC1333: \citealt{doi2020,doi2021}; B1: \citealt{coude2019}).
	The gas kinematics at molecular cloud scales were reported for NGC1333 \citep{hacar2017,chen2020}.
	The multiplicity of all Perseus protostars has been characterized down to 15 AU projected separations \citep{tobin2016}.
	Parameters derived from SEDs, including \textit{Herschel} Space Observatory PACS fluxes, are available for Perseus protostellar sources \citep{murillo2016}.
	
	The maps presented in this work cover five subregions: NGC1333, L1448, L1455, B1, and IC348 (Fig.\ref{fig:TKmap_N2H}).
	These maps cover clustered (NGC1333), non-clustered (L1448, L1455, B1, IC348), externally irradiated (IC348, NGC1333, L1448), and relatively isolated (B1, L1455) regions.
	The protostellar population sampled includes objects from Class 0 to Class II, obtained from the VANDAM survey \citep{tobin2016}, as well as starless cores obtained from \citet{hatchell2007}.
	The multiplicity of these objects comprise bonafide single protostars, close (here defined as projected separations $<$7$\arcsec$) and wide (defined in this work as projected separations $\geq$ 7$\arcsec$) multiple protostellar systems. 
	The projected separation classification is based on the systems that can be resolved with \textit{Herschel PACS} observations \citep{murillo2016}, and makes sense for the Nobeyama maps with  pixel size of 6$\arcsec$. 
	The protostellar systems and starless cores are listed in Table~\ref{tab:source} and~\ref{tab:starless}, respectively.

	\section{Observations}
	\label{sec:observations}
	
	\subsection{Nobeyama 45m Radio Observatory (NRO)}
	\label{subsec:nro}
	Observations of Perseus with the Nobeyama 45m Radio Telescope (NRO) were done using the FOur-beam REceiver System on the 45-m Telescope (FOREST, \citealt{minamidani2016}) frontend, and the Spectral Analysis Machine for the 45-m Telescope (SAM45, \citealt{kamazaki2012}) backend in the project CG201020 (PI: N. M. Murillo).
	Observations were carried out between January and March 2021.
	OTF mapping mode was used to make 162$\arcsec \times$162$\arcsec$ maps with a 6$\arcsec$ grid to maximize coverage of the sample (Table~\ref{tab:source}) and time constraints.
	Smaller maps allowed an efficient use of the observing time ($\sim$45 minutes per map), and streamlining data reduction.
	Due to time constraints, the IRAS 03282+3035 is smaller (about 120$\arcsec \times$130$\arcsec$), and B1-a was observed in only one frequency setup, as the source can be detected in adjacent maps.

	Two spectral setups were used to target eight molecular species: \ce{HCN} $J$=1--0, \ce{HCO+} $J$=1--0, \ce{HNC} $J$=1--0, \ce{HC3N} $J$=10--9, \ce{c-C3H2} $J$=7$_{6,2}$--7$_{5,3}$, \ce{N2H+} $J$=1--0, \ce{^{13}CO} $J$=1--0 and \ce{C^{18}O} $J$=1--0.
	The targeted \ce{c-C3H2} transition ($E_{up}$ = 87.45 K)  was not detected.
	Cyanoacetylene (\ce{HC3N} $J$=10--9) will be discussed in a future work.
	The spectral windows have a channel resolution of 30.52 kHz (0.1 km~s$^{-1}$ per channel at 90.6 GHz) and a bandwidth of 125 MHz.
	Noise levels range between 0.3 and 3.3 K, with an angular resolution between 18.7$\arcsec$ (\ce{HCN}) and 15.1$\arcsec$ (\ce{C^{18}O}).
	Beam efficiencies $\eta_{\rm mb}$ of about 0.5 were adopted based on the measurements by NRO for the 2021-2022 observing season. 
	
	Standard data calibration and imaging for each datacube was done with the NOSTAR software package provided by NRO.
	Contiguous individual maps were combined into mosaics for NGC1333, L1448, L1455, B1, and IC348 after determining the image center, bottom left corner, and top right corner coordinates.
	The maps for L1455 Per-emb 25 (hereafter Per25), IRAS 03292+3039 (hereafter IRAS03292), and IRAS 03282+3035 (hereafter IRAS03282) are shown individually given their locations in Perseus.
	All maps have been clipped to only show values above 3$\sigma$ or 5$\sigma$. 
	Further analysis was carried out using CASA \citep{casa2022} and self-written python scripts using the Astropy \citep{astropy:2013,astropy:2018} and PySpecKit \citep{ginsburg2011} libraries.
	Intensity integrated maps and respective uncertainties were obtained with the bettermoments python library \citep{teague2018,teague2019}. 
	See Appendix~\ref{ap:moment} for details.
	
	\subsection{Atacama Pathfinder EXperiment (APEX)}
	\label{subsec:apex}
	Single-pointing observations of \ce{HNC} $J$=4--3 toward 37 protostellar systems were performed with the Atacama Pathfinder EXperiment (APEX, \citealt{gusten2006}) using the APEX-2 and First Light APEX Submillimeter Heterodyne (FLASH$^{\rm +}$) receivers. 
	Some of the protostellar systems covered by the NRO maps were not observed with APEX (e.g., NGC1333 IRAS4).
	Ten protostellar systems were observed with APEX-2 \citep{murillo2018}. 
	The \ce{HNC} $J$=4--3 spectra have native channel widths of 0.1 km~s$^{-1}$, and binned to 0.4 km~s$^{-1}$ to increase the signal-to-noise ratio (S/N).
	Typical noise levels range between 20 to 100 mK for \ce{HNC} $J$=4--3 with APEX-2, for a channel width of 0.4 km s$^{-1}$, and a half-power beam width (HPBW) of 18$\arcsec$.
	FLASH$^{\rm +}$ observations (Project ID: O-0104.F-9307B) of 27 protostellar systems were done on six dates: 11, 15--18, 20 October 2019.
	The spectral setup has a central frequency of 361.16978 GHz and a bandwidth of 4 GHz.
	Typical noise levels ranged between 25 and 145 mK for a channel width of 0.4 km s$^{-1}$, and a HPBW of 18$\arcsec$.
	A beam efficiency of $\eta_{\rm mb}$ = 0.73 was adopted for both instruments.
	
	\subsection{Data from literature}
	\label{subsec:litdata}
    Observations of \ce{HNC} $J$=4--3  with the James Clerk Maxwell Telescope (JCMT) toward NGC1333 IRAS4 were obtained from the tabulated values in \citet{koumpia2016}. JCMT observations have an angular resolution of 15$\arcsec$, and thus a beam dilution correction of 0.69 is applied to the values from \citet{koumpia2016} in order to compare with our APEX data.
	The Perseus dust temperature map and its corresponding uncertainty map \citep{zari2016} were downloaded from the CDS archive\footnote{\url{https://cdsarc.cds.unistra.fr/viz-bin/cat/J/A+A/587/A106}}. 
	The dust temperature is derived from the combined \textit{Herschel} and \textit{Planck} maps of Perseus using SED fitting with a modified blackbody assumption.
	The Green Bank Ammonia Survey (GAS) map of NGC1333 \citep{friesen2017} was obtained from the Dataverse archive, and is derived from fitting of \ce{NH3} line emission.
	Information on molecular outflows driven by the protostars included in the NRO maps is obtained from the MASSES survey \citep{stephens2017}.
	
	\begin{figure*} 
		\centering
		\includegraphics[width=0.97\linewidth]{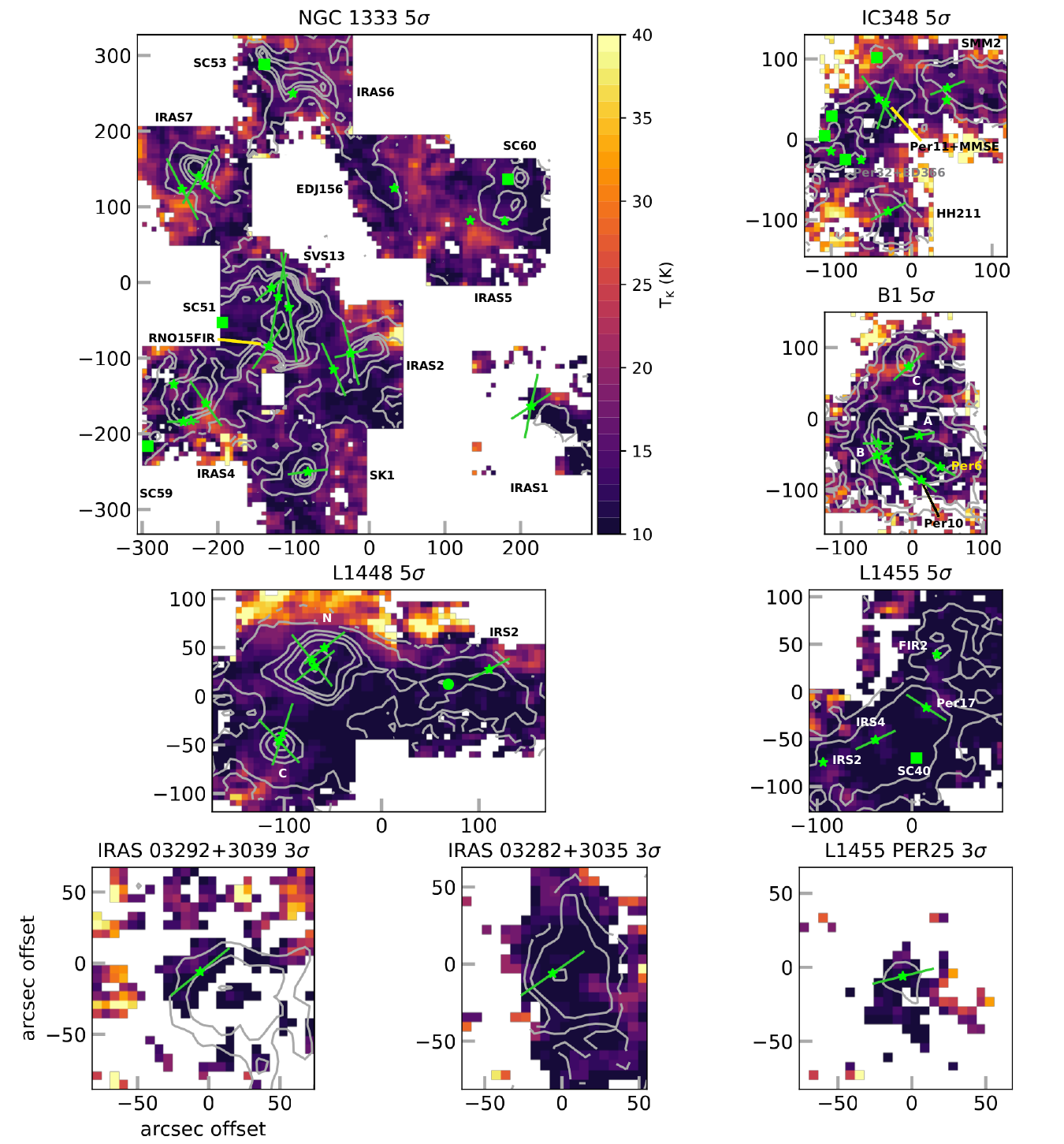}
		\caption{Gas kinetic temperature map derived from the $I$(\ce{HCN})/$I$(\ce{HNC}) $J$=1--0 ratio (color scale, same as Fig.~\ref{fig:TKmap}) overlaid with the integrated intensity map of diazenylium \ce{N2H+} $J$=1--0 (gray contours in steps of 3, 5, 10, 15, 20, 25 and 30 K~km~s$^{-1}$, see also Fig.~\ref{fig:N2Hmom0}) for each region observed. All maps are shown with the same color scale range for comparison. The $\sigma$ value above each panel indicates the cutoff value for both maps. Star symbols mark the positions of protostellar sources, and squares mark the locations of starless cores. The filled circle symbol marks the position of L1448 IRS2E whose nature is debated. Straight lines indicate the outflow directions for the protostellar systems included in the MASSES survey \citep{stephens2017}.}
	    \label{fig:TKmap_N2H}
    \end{figure*}

    \begin{figure*}
    	\centering
    	\includegraphics[width=0.89\linewidth]{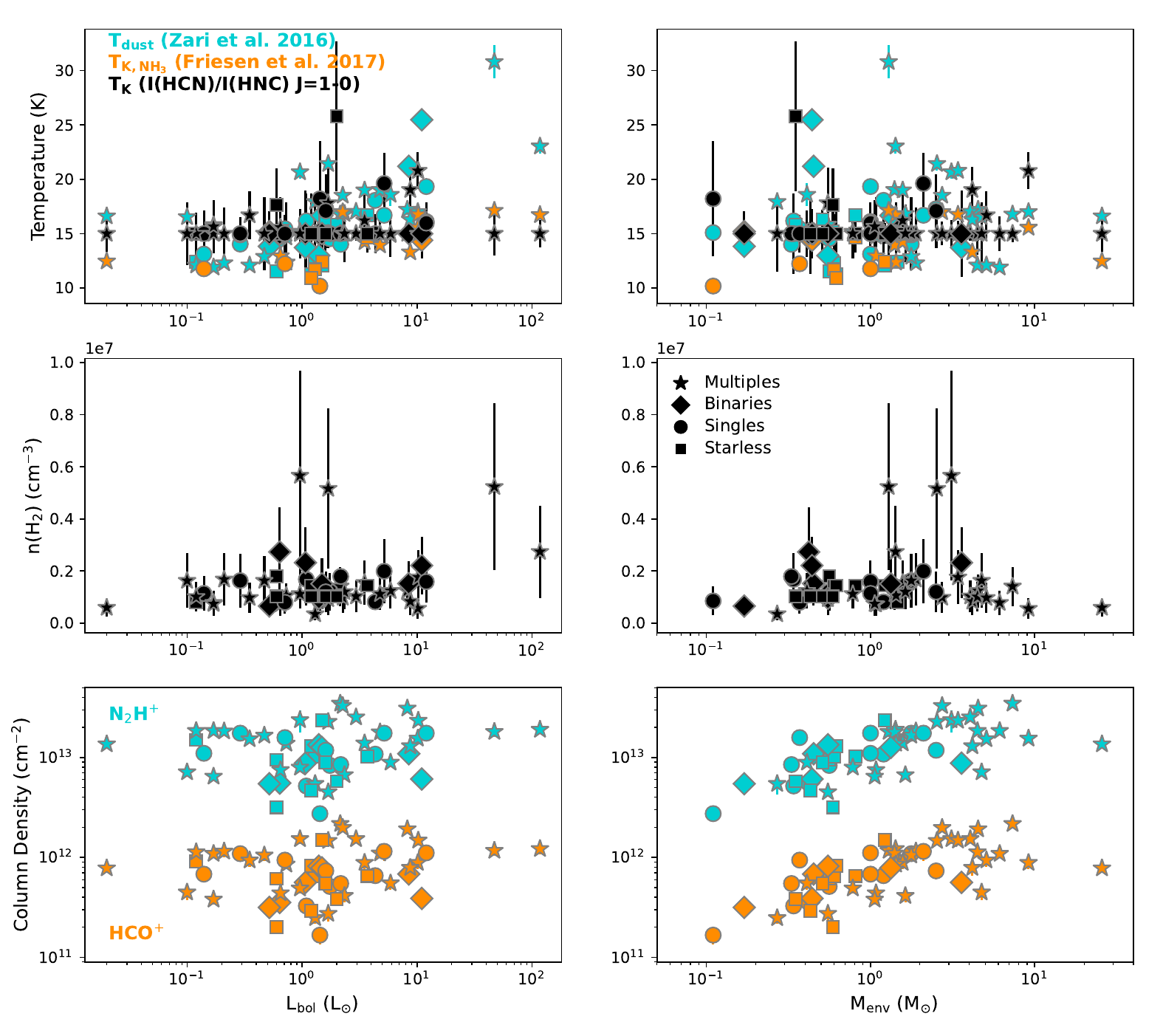}
    	\caption{Gas and dust Temperature (\textit{first row}), derived \ce{H2} volumetric density from the \ce{HNC} $J$=4--3/$J$=1--0 ratio (\textit{second row}), and derived \ce{N2H+} $J$=1--0 and \ce{HCO+} $J$=1--0 column density (\textit{third row}) versus source bolometric luminosity $L_{\rm bol}$ (\textit{left column}), and envelope mass $M_{\rm env}$ (\textit{right column}). In the bottom row, the error bars are about the size of the plotted points. The starless cores (squares), HH211, and L1445 IRS2 adopt the average \ce{H2} volume density for their respective region (see Table~\ref{tab:derivednNM}).}
    	\label{fig:pointplots}
    \end{figure*}
	
	\begin{figure*}
		\centering
		\includegraphics[width=0.89\linewidth]{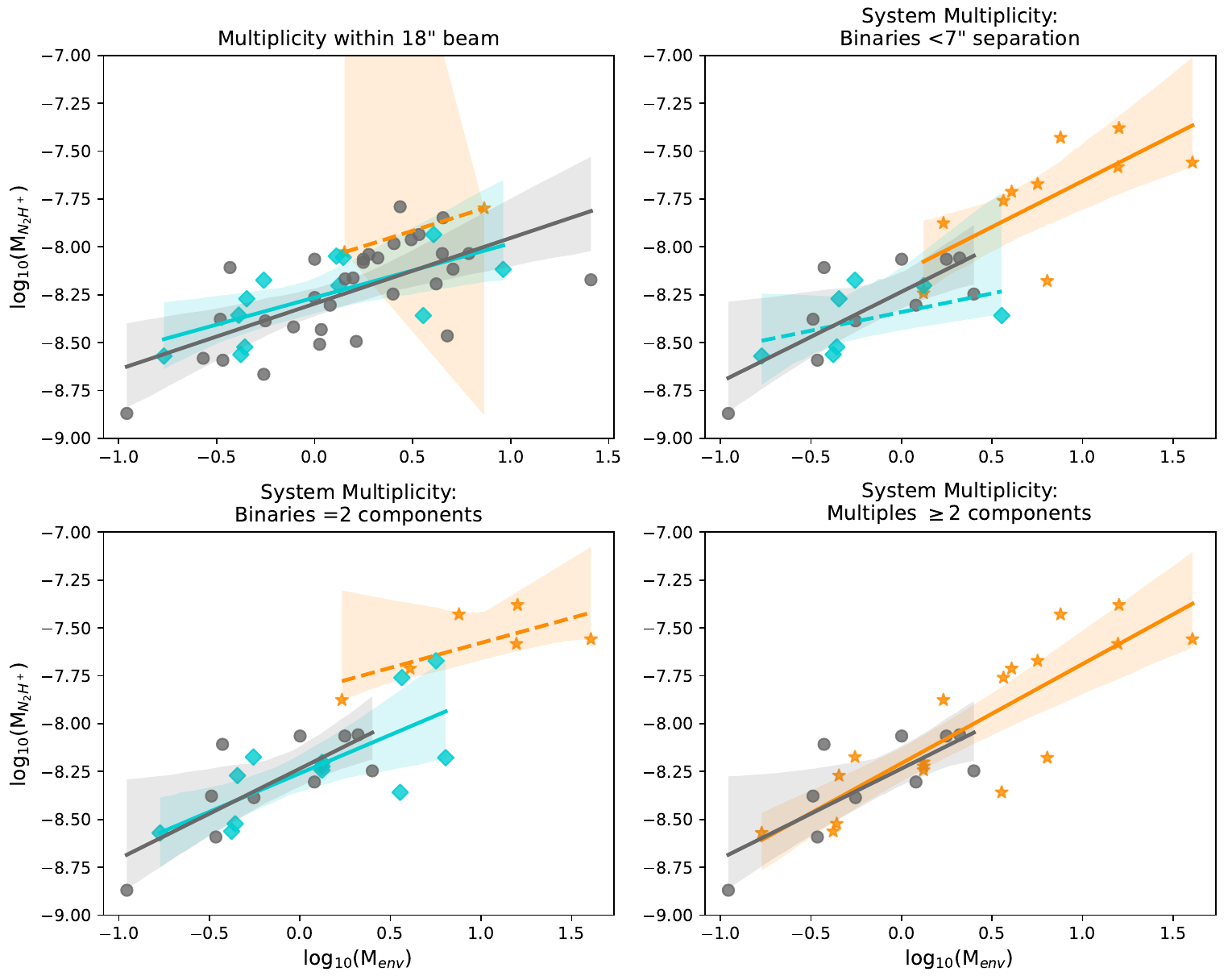}
		\caption{Relations between \ce{N2H+} gas mass, M$_{gas}$(\ce{N2H+}, and envelope dust mass, $M_{env}$, for the sample in this work. The orange stars show multiple systems, the cyan diamonds show binary systems, and the gray circles show single protostellar systems. Each panel represents one of four ways of grouping the sample presented in this work, and their corresponding correlations. The lines and shaded areas show the linear regression for the data with the corresponding color. Solid lines indicate statistically significant correlations (Pearson $r$ and Spearman $\rho$ p-values$<$0.05), while dashed lines show subsamples with p-values$>$0.05. See Section~\ref{subsec:stats} for discussion on the figure.} 
		\label{fig:massColDen}
	\end{figure*}
	
	\section{Results}
	\label{sec:results}

	\subsection{Nobeyama maps}
	\label{subsec:NROmaps}
	
	In this paper we focus on emission from \ce{HCN}, \ce{HCO+}, \ce{HNC}, \ce{N2H+}, \ce{^{13}CO}, and \ce{C^{18}O}, all in the $J$=1--0 transition.
	Detailed description and integrated intensity maps of each molecular species are given in Appendix~\ref{ap:moment}.
	Source positions \citep{hatchell2007,tobin2016} and known outflow directions \citep{stephens2017} are plotted in all maps.
	The spatial gas distributions of \ce{HCN}, \ce{HCO+}, \ce{HNC} and \ce{N2H+} show bright and extended emission around higher order multiple protostellar systems ($>$2 components), and weaker more compact emission around isolated systems or close binaries and single protostars (Fig.~\ref{fig:HCNmom0}--\ref{fig:N2Hmom0}).
	The strongest emission of these four species is mainly located in NGC1333 and L1448.
	The spatial distribution of the two \ce{CO} isotopologues, \ce{^{13}CO} and \ce{C^{18}O}, does not match any of the other molecular lines.
	The main core of the B1 region presents the brightest \ce{C^{18}O} emission, but \ce{HCO+} and \ce{^{13}CO} are weaker.

	\subsection{Single pointing observations of \ce{HNC}} 
	\label{subsec:APEXHNC}
	Single pointing observations with APEX detected \ce{HNC} $J$=4--3 toward all sources in the sample with signal-to-noise ratio (S/N) of 4 or higher.
	The APEX observations did not include the NGC1333 IRAS4 mini-cluster, HH211, L1455 IRS2 (Table~\ref{tab:source}) or the starless cores (Table~\ref{tab:starless}).
	Peak values range from 0.17 to 3.3 K, with an average noise level of 50 mK.
	The strongest \ce{HNC} $J$=4--3 emission (S/N $>$ 48) is present toward L1448 C, NGC1333 IRAS6, NGC1333 SVS13, and B1-C. 
	The weakest emission (4 $<$ S/N $<$ 6) is present toward NGC1333 EDJ2009156, IC348 EDJ2009366, and NGC1333 IRAS5 Per63.
	The average line width for the \ce{HNC} $J$=4--3 transition is 1 km~s$^{-1}$, with a range of 0.5 to 1.9 km~s$^{-1}$.
	 
	Spectra from the \ce{HNC} $J$=1--0 NRO maps was extracted from 18$\arcsec$ circular regions centered on the positions listed in Table~\ref{tab:source} to match the beam size of the APEX observations.
	Peak brightness varies from 1.4 to 7.2 K, with an average noise level of 144 mK.
	The weakest emission is found toward two sources, L1455 Per25 (S/N = 8) and NGC1333 IRAS4C (S/N = 9). 
	All other source positions have emission with S/N $>$ 11.
	The $J$=1--0 transition presents an average line width of 1.5 km~s$^{-1}$ with a range of 0.9 to 2.3 km~s$^{-1}$, slightly wider than the 4--3 transition.

	\section{Analysis}
	\label{sec:analysis}
	The physical parameters analyzed in this section are mean values obtained from within a circular region of 18$\arcsec$ in diameter centered on the sources listed in Tables~\ref{tab:source} and \ref{tab:starless}, and are referred to as the physical parameter of the source for simplicity.
	Hence the reported values consider envelope scales ($\sim$5000 AU), and not the entire molecular cloud.
	
	\subsection{Envelope dust mass}
	\label{subsec:dustmass}
	Envelope dust mass is calculated based on the relation from \citet{jorgensen2009}, namely
	\begin{equation}
		\label{eq:dustmass}
		M_{\rm env} = 0.44 M_{\odot}~\left(\frac{L_{\rm bol}}{1~L_{\odot}}\right)^{-0.36}~\left(\frac{S_{850\mu m}}{1~{\rm Jy~beam}^{-1}}\right)^{1.2}~\left(\frac{d}{125~{\rm pc}}\right)^{1.2}
	\end{equation}
	where envelope dust mass, $M_{\rm env}$, is calculated from the 850 $\mu$m dust continuum peak intensity $S_{\rm 850\mu m}$ within the 18$\arcsec$ beam ($\sim$5000 AU), bolometric luminosity $L_{\rm bol}$ and distance $d$.
	The dust continuum map at 850$\mu$m is obtained from the COMPLETE survey map of Perseus \citep{kirk2006}, and the peaks are obtained from within circular regions of 18$\arcsec$ centered on the sources in our sample (Table~\ref{tab:source} or \ref{tab:starless}).
	The distance $d$ for each subregion is taken from \citet{zucker2018}, adopting 288 pc for L1448, 299 pc for NGC1333 and L1455, 301 pc for B1, and 295 pc for IC348.
	
	\subsection{Gas kinetic temperature from $I$(\ce{HCN})/$I$(\ce{HNC})}
	\label{subsec:gasT}

	The clipped (3 or 5$\sigma$) integrated intensity \ce{HCN} and \ce{HNC} maps were used to make the $I$(\ce{HCN})/$I$(\ce{HNC}) $J$=1--0 ratio map, which was then converted to kinetic temperature using the equations from \citet{hacar2020}
	\begin{equation}
		\label{eg:hcnhnctemp}
		T_{K} [K] = 10 \times \left[\dfrac{I(\ce{HCN})}{I(\ce{HNC})}\right] ~if~ \left[\dfrac{I(\ce{HCN})}{I(\ce{HNC})}\right] \leq 4.
	\end{equation}
	We note that the $I$(\ce{HCN})/$I$(\ce{HNC}) $J$=1--0 maps are insensitive to temperatures $<$15 K and $>$40 K.
	Hence values beyond these limits are arbitrarily set to the 15 and 40 K, respectively, for subsequent calculations of physical parameters.
	The $I$(\ce{HCN})/$I$(\ce{HNC}) $J$=1--0 ratio uncertainty map was generated, assuming zero covariance, with the relation
	\begin{equation}
		\label{eg:error}
		\sigma^{2}_{T_{K}} = T_{K}^{2} \left[(\dfrac{\sigma_{I(\ce{HCN})}}{I(\ce{HCN})})^{2} + (\dfrac{\sigma_{I(\ce{HNC})}}{I(\ce{HNC})})^{2}\right].
	\end{equation}
	Errors were extracted from the uncertainty map using circular regions of 18$\arcsec$.
	
	Figure~\ref{fig:TKmap_N2H} shows the derived gas kinetic temperature $T_{K}$ map. 
	The integrated intensity map of \ce{N2H+} $J$=1--0, tracing cold dense gas, is overlaid on the $T_{K}$ map.
	The subregion mean $I$(\ce{HCN})/$I$(\ce{HNC}) ratio was calculated considering all the pixels above 5$\sigma$, resulting in mean ratios of  1.5$\pm$0.6 for NGC1333, 1.5$\pm$0.8 for L1448, 2.0$\pm$0.9 for IC348, 1.7$\pm$1.1 for B1, 1.1$\pm$0.4 for L1455, 2.5$\pm$1.1 for IRAS03292, 1.2$\pm$0.4 for IRAS03282, and 1.4$\pm$0.4 for L1455 Per25.
	The $T_{K}$ map without overlay is shown in Figure~\ref{fig:TKmap}, while Figures~\ref{fig:HCNmom0} to \ref{fig:13COmom0} show the $T_{K}$ map contours overlaid on the integrated intensity maps of the other molecules for comparison. Table~\ref{tab:derivedT} lists the mean $I$(\ce{HCN})/$I$(\ce{HNC}) $J$=1--0 ratios, uncertainty and adopted $T_{K}$ within 18$\arcsec$ circular regions centered on the sources in the sample.
	
	The mean $T_{K}$ versus $L_{\rm bol}$ (Table~\ref{tab:source}) and $M_{\rm env}$ (Table~\ref{tab:derivednNM}) for each source are plotted in Fig.~\ref{fig:pointplots} along with the dust temperature $T_{dust}$ from combined \textit{Herschel} and \textit{Planck} maps (36$\arcsec$ within the \textit{Herschel} map area: \citealt{zari2016}).
	For sources in NGC1333, the \ce{NH3} gas kinetic temperature $T_{K,\ce{NH3}}$ from the Green Bank Ammonia Survey is included (GAS, 32$\arcsec$ beam: \citealt{friesen2017}) for comparison.
	The values were all obtained from within a circular region with 18$\arcsec$ diameter, and are listed in Table~\ref{tab:derivedT}.

	\subsection{\ce{H2} density from \ce{HNC} $J$=4-3 / $J$=1-0}
	\label{subsec:H2density}
	
	Considering the kinetic temperature constraints derived from $I$(\ce{HCN})/$I$(\ce{HNC}) and the availability of \ce{HNC} $J$=1--0 and 4--3 data with the same angular resolution, the \ce{H2} volumetric density n(\ce{H2}) can be derived based on observational data and radiative transfer models.
	Non-LTE excitation and radiative transfer calculations using RADEX \citep{vandertak2007} were performed to model the \ce{HNC} $J$=4--3 / $J$=1--0 ratio with respect to \ce{H2} volumetric density n(\ce{H2}) and gas kinetic temperature.
	Using these models, the observed \ce{HNC} $J$=4--3 / $J$=1--0 ratio (assuming both transitions arise from the same gas) together with $T_{K}$ enable us to obtain mean n(\ce{H2}) within 5000 AU regions.
	APEX and NRO observations have the same angular resolution ($\sim$18$\arcsec$), thus no beam dilution correction is needed. 
	A beam dilution correction (0.69) is applied to the JCMT data (15$\arcsec$ beam).
	
	The calculations adopted a N(\ce{HNC}) = 10$^{13}$ cm$^{-2}$ and a line width of 1 km~s$^{-1}$ (See Section~\ref{subsec:APEXHNC}), as these parameters best reproduce the observed \ce{HNC} peak emission.
	The modeled n(\ce{H2}) values were selected using the $T_{K}$ and \ce{HNC} $J$=4--3 / $J$=1--0 ratio constraints, with corresponding uncertainties, for each source.
	The obtained n(\ce{H2}) values for each source were then averaged and the standard derivation calculated to obtain the adopted value and corresponding uncertainty per source.
	The adopted n(\ce{H2}) with derived $T_{K}$ and corresponding uncertainties are  listed in Table~\ref{tab:derivednNM}.
	An order of magnitude higher N(\ce{HNC}) results in n(\ce{H2}) near or below the critical density of the \ce{HNC} 1--0 transition ($\sim$2$\times$10$^{5}$ cm$^{-3}$ between 5 and 40 K).
	If N(\ce{HNC}) is an order of magnitude lower than adopted, n(\ce{H2}) varies between 7$\times$10$^{6}$ and 2$\times$10$^{9}$ cm$^{-3}$, which are typical densities of low-mass protostellar envelope models at R $<$ 1000 AU (e.g., \citealt{jorgensen2005,vanthoff2018}).
	Broader line widths of 1.5 and 2.0 km~s$^{-1}$ (See Section~\ref{subsec:APEXHNC}) varies the calculated mean n(\ce{H2}) by less than 10\%, except for L1448 C which varies as much as 60\%.
	The n(\ce{H2}) derived from the \ce{HNC} $J$=4--3 / $J$=1--0 ratio versus $L_{\rm bol}$, and $M_{\rm env}$ are shown in Fig.~\ref{fig:pointplots}.
	
	\subsection{Column density and gas mass from \ce{N2H+} $J$=1--0 and \ce{HCO+} $J$=1--0}
	\label{subsec:colden_mass}
	
	The n(\ce{H2}) (Section~\ref{subsec:H2density}), $T_{K}$ (Section~\ref{subsec:gasT}), peak brightness temperature and line width are used to derive the column density of \ce{N2H+} (N(\ce{N2H+})) and \ce{HCO+} (N(\ce{HCO+})) using RADEX.
	An average of each subregion was calculated and adopted as the n(\ce{H2}) for sources without \ce{HNC} $J$=4--3 data.
	Uncertainties for the column densities were calculated by using the lower and upper limits of n(\ce{H2}) and $T_{K}$ in the RADEX calculations, and then taking the average and standard deviation of all three values.
	Changes of an order of magnitude in n(\ce{H2}) do not significantly affect the derived N(\ce{N2H+}) and N(\ce{HCO+}).
	To test whether using $T_{K}$, $T_{K,\ce{NH3}}$ or $T_{dust}$ affected the calculations, N(\ce{N2H+}) and N(\ce{HCO+}) toward NGC1333 were derived using the available temperatures.
	The results show that N(\ce{N2H+}) and N(\ce{HCO+}) are the same regardless of the temperature used, indicating that the calculations are sensitive to the line peak brightness temperature and line width of the corresponding emission.
	
	The weaker and optically thin component of \ce{N2H+} at 93.1739 GHz is used to derive N(\ce{N2H+}), as it is optically thin and less prone to blending.
	The bettermoments python library \citep{teague2018,teague2019} is used to perform a Gaussian fit in order to obtain peak brightness temperature and line width \ce{N2H+} and \ce{HCO+} maps.
	The mean peak brightness temperatures and line widths are then extracted for the source positions within 18$\arcsec$ regions.

	The gas mass in M$_{\odot}$ for \ce{N2H+} (M$_{gas}$(\ce{N2H+})) and \ce{HCO+} (M$_{gas}$(\ce{HCO+})) within 18$\arcsec$ diameter regions  is derived with
		\begin{equation}
			\label{eg:mass}
			M_{gas} = \frac{M_{r}~(g~mol^{-1}) N~(cm^{-2})~A~(cm^2)}{N_{A}~(mol^{-1})~M_{sun,g} ~(g)} ~~M_{\odot}
		\end{equation}
	where $M_{r}$ is the mean molecular weight (29.022 and 29.018 g~mol$^{-1}$ for \ce{N2H+} and \ce{HCO+}, respectively), $N_{A}$ is Avogadro's constant, $N$ is the derived gas column density, $A$ is the area, and $M_{sun, g}$ is the solar mass in grams.  
	The uncertainty is obtained with the same equation, but with N being the uncertainty of the derived column density.
	Distances used are the same as noted in Section~\ref{subsec:dustmass}.

    Derived N(\ce{N2H+}) and N(\ce{HCO+}) are plotted in Figure~\ref{fig:pointplots} versus $L_{\rm bol}$, and $M_{\rm env}$. 	
    The \ce{HHCO+} column density is lower by a factor of $\sim$16 relative to the \ce{N2H+} column density, and both follow the same trend (Fig.~\ref{fig:pointplots}, Table~\ref{tab:derivednNM}).
    Hence, M$_{gas}$(\ce{N2H+})  is shown in subsequent plots to illustrate the relation between gas mass, dust mass, and multiplicity in Figure~\ref{fig:massColDen}, with M$_{gas}$(\ce{HCO+}) following the same trend.
	
	Within the regions probed (18$\arcsec$, ~5000 AU), \ce{N2H+} traces cold ($\leq$20 -- 30 K) dense (typically $\geq$ 10$^{5}$ cm$^{-3}$) gas, while \ce{HCO+} traces warmer gas ($>$20 -- 30 K).
	This is due to the chemical formation and destruction pathways of both molecules being related to the presence of \ce{CO} (e.g., \citealt{murillo2022b} ).
	\ce{N2H+} is present in the gas phase when \ce{CO} is frozen out onto the dust grains (temperatures below 20 -- 30 K). 
	Meanwhile, \ce{HCO+} formation requires \ce{CO} to be in the gas phase and is destroyed by molecules like water, thus \ce{HCO+} is present in gas at temperatures above 20 -- 30 K and up to 100 K.

	\subsection{Statistics}
	\label{subsec:stats}
	Three statistical tests are performed to examine the relations between physical parameters and multiplicity.
	The number of protostellar components is taken from \citet{tobin2016}, and updated based on \citet{reynolds2023}.
	A 2-sample Anderson-Darling (AD)\footnote{The Scipy function used, anderson\_ksamp, requires a sample size of 2 or more, as noted in the Scipy API \url{https://docs.scipy.org/doc/scipy/reference/generated/scipy.stats.anderson_ksamp.html}. The calculated statistic of the test increases with decreasing sample size.} test assuming normal distribution is used to check whether parameters share the same parent distribution.
	An AD statistic $<$1.961 and $p-value >$ 0.05 indicates that both samples arise from the same parent distribution. 
	The Pearson linear correlation coefficient ($r$) and Spearman's rank correlation coefficient ($\rho$) are used to find correlated parameters. 
	The values of $r$ and $\rho$ range from -1 to 1, for negative slope to positive slope correlations, respectively, with values closer to 1 (absolute value) indicating stronger correlations. 
	Statistically significant correlations are identified by $p-values <$ 0.05.
	Linear regression is used to visualize the correlations between physical parameters.
	All tests are done using the statistical functions in the Python library Scipy.
	Five classification schemes, excluding starless cores, are used for the statistical tests:
	\begin{enumerate}
		\item Full sample, parameters within 18$\arcsec$ diameter regions. The sample size is 46 sources.
		\item Multiplicity within 18$\arcsec$ diameter regions: each continuum peak is a separate system (e.g., SVS13A, B and C are three individual systems). This scheme considers that sources with separations larger than 5000 AU are not gravitationally bound. Bin sizes: Multiples = 2 systems; Binaries = 12 systems; Singles = 32 systems.
		\item System multiplicity with binaries having separations $<$7$\arcsec$. This scheme considers core vs. disk fragmentation scenarios. Bin sizes: Multiples = 11 systems; Binaries = 7 systems; Singles = 11 systems.
		\item System multiplicity following the classification of singles (1 component), binaries (2 components), and higher order multiples ($>$2 components). Bin sizes: Multiples = 6 systems; Binaries = 12 systems; Singles = 11 systems.
		\item System multiplicity where multiples are systems with 2 or more components, examining the scenario where all multiple systems share formation mechanisms. Bin sizes: Multiples = 18 systems; Singles = 11 systems.
	\end{enumerate}
	Systems are considered to be all the protostars within a common core, and thus M$_{gas}$(\ce{N2H+}), M$_{gas}$(\ce{HCO+}) and M$_{env}$ of a system is the sum of the masses of each of its components.
	The parameters such as $T_{K,\ce{NH3}}$, $T_{dust}$, $T_{K}$, n(\ce{H2}), N(\ce{N2H+}) and N(\ce{HCO+}) are averaged from all the components.
	
	\subsubsection{Temperature}
	For the full sample, the AD test indicates that $T_{K,\ce{NH3}}$ \citep{friesen2017}, $T_{dust}$ \citep{zari2016}, and $T_{K}$ (this work), do not share a common parent distribution.
	If the lower limit of 15 K is not applied to the $I$(\ce{HCN})/$I$(\ce{HNC}) $J$=1--0 ratio (see Section~\ref{subsec:gasT}), then  $T_{K,\ce{NH3}}$ and $T_{K}$ do share the same parent distribution ($p-value$ = 0.71). However, in this work the $T_{K}$ with the 15 K lower limit is adopted.
	Using the full sample, $T_{dust}$ shows a correlation with $L_{\rm bol}$ ($r \sim$0.7, $\rho \sim$0.7, $p-value <<$0.05), but not with mass.
	Given the method used to obtain $T_{dust}$ (see Section~\ref{subsec:litdata}, \citealt{zari2016}) and $L_{bol}$ \citep{murillo2016}, the correlation is expected as the quantities are not independent.
	A weak correlation between number of protostellar components and $T_{dust}$ is found, but this may be skewed due to the source with the highest luminosity also has the highest number of components (SVS13A, 117 L$_{\odot}$, 4 protostars), while single sources have $L_{bol} \leq$ 12 L$_{\odot}$.
	On the other hand, $T_{K}$ shows no correlation to $L_{bol}$, mass or number of protostellar components, for any of the five classification schemes used.
	
	For NGC1333, $T_{K,\ce{NH3}}$ shows a correlation to bolometric luminosity ($r \sim$0.6, $\rho \sim$0.6, $p-value <$0.006), N(\ce{N2H+}), N(\ce{HCO+}), and the corresponding gas masses ($r \sim$0.6, $\rho \sim$0.6, $p-value <$0.006). 
	In the method used to derive $T_{K,\ce{NH3}}$, N(\ce{NH3}) is implicitly dependent on gas kinetic temperature \citep{friesen2017}.
	This result suggests that N(\ce{N2H+}) and N(\ce{NH3}) are correlated, confirming previous work \citep{johnstone2010}.
	
	\subsubsection{Density}
	Considering the full sample, the AD test indicates that N(\ce{N2H+}) and N(\ce{HCO+}) are not drawn from the same parent distribution.
	The n(\ce{H2}) shows a weak correlation to dust temperature ($r \sim$0.4, $\rho \sim$0.3, $p-value <$0.04), and an even weaker linear but not ranked correlation to $L_{\rm bol}$ ($r \sim$0.3, $p-value \sim$0.03).
	No correlation is found between $T_{K}$ and the derived n(\ce{H2}) from this work, or between the number of protostellar components and n(\ce{H2}).
	A visual representation of n(\ce{H2}) versus $T_{K}$ and dense cold gas (\ce{N2H+}) maps is shown in Fig.~\ref{fig:TKmap_H2density}.
	
	In contrast, N(\ce{N2H+}) and N(\ce{HCO+}) show no correlation to $T_{dust}$, $L_{\rm bol}$, $T_{K}$, $M_{\rm env}$, or number of protostellar components.
	As expected from the method used in this work, N(\ce{N2H+}) and N(\ce{HCO+}) show a ``perfect" correlation to  M$_{tot}$(\ce{N2H+}) and M$_{tot}$(\ce{HCO+}) ($r \sim$1.0, $\rho \sim$1.0, with very small $p-values$).
	Hence, the correlation tests are not run on the subsamples from classification schemes 2 -- 5 for N(\ce{N2H+}) and N(\ce{HCO+}), as column densities would follow the same trends as gas mass.
	
	\subsubsection{Mass} 
	For all five classification schemes, M$_{gas}$(\ce{N2H+}), M$_{gas}$(\ce{HCO+}), $M_{\rm env}$ are not consistent with being drawn from the same parent distribution.
	The AD test does show that the same mass parameter between binary and single protostellar subsamples in classification schemes 2 -- 4 are drawn from the same parent distribution, with $p-values>$0.4 for gas masses and $p-values>$0.3 for dust masses.
	This shared distribution is visible in the $M_{\rm env}$ vs M$_{gas}$(\ce{N2H+})  plots (Figs.~\ref{fig:massColDen} and \ref{fig:numcomp}).
	The multiples subsample in classification schemes 3 and 4 does not share parent distributions with binary and single subsamples, for any of the tested parameters.
	For classification scheme 5, M$_{gas}$(\ce{N2H+}) and M$_{gas}$(\ce{HCO+}) for the multiple and single populations are found to be drawn from the same parent distribution (AD test $p-value~\sim$0.1), but the $M_{\rm env}$ does not show the same result (AD test $p-value~\sim$0.03).
	
	For the full sample, $M_{\rm env}$ shows a positive correlation with M$_{gas}$(\ce{N2H+}) and M$_{gas}$(\ce{HCO+}), both cases having $r \sim$0.7, $\rho \sim$0.7, with $p-value <<$0.05.
	Concerning classification scheme 2, $M_{\rm env}$ shows positive correlation with M$_{gas}$(\ce{N2H+}) and M$_{gas}$(\ce{HCO+}) for the binary  ($r \sim$0.7, $\rho \sim$0.7, $p-value <$0.02) and single ($r \sim$0.6, $\rho \sim$0.6, $p-value <<$0.01) subsamples (Fig.~\ref{fig:massColDen}, first panel). 
	The multiples subsample only contained two sources, and thus no significant statistical analysis could be performed.
	
	For classification scheme 3 (binaries with separations of $<$7$\arcsec$; Fig.~\ref{fig:massColDen}, second panel), strong positive correlations are found between dust and gas masses for singles ($r \sim$0.6, $\rho \sim$0.6, $p-value <<$0.05) and multiples ($r \sim$0.7, $\rho \sim$0.8, $p-value <<$0.05).
	The binary subsample does not show correlation, according to the Pearson $r$ and Spearman $\rho$ tests.
	This may be due to the small sample size of the binaries (7 sources).
	
	In classification scheme 4 (binaries have two components; Fig.~\ref{fig:massColDen}, third panel), $M_{\rm env}$ shows positive correlation to gas masses for binaries ($r \sim$0.7, $\rho \sim$0.8, $p-value <$0.01) and singles ($r \sim$0.8, $\rho \sim$0.7, $p-value <$0.03).
	The multiples subsample has a small sample size (6 sources), and thus does not show a statistically significant correlation.
	Classification scheme 5 (multiples have $\geq$2 components; Fig.~\ref{fig:massColDen}, fourth panel) shows a correlation between $M_{\rm env}$ and gas mass for multiples ($r \sim$0.8, $\rho \sim$0.9, $p-value <<$0.001) and singles ($r \sim$0.8, $\rho \sim$0.7, $p-value <$0.03).

	\subsubsection{Number of components} 
	For the full sample and classification scheme 5, the number of protostellar components shows a positive correlation with dust and gas masses ($r \sim$0.6, $\rho \sim$0.6, with $p-value <<$0.05). 
	For classification scheme 3, a correlation is found for the multiples subsample between gas masses and number of protostellar components  ($r \sim$0.8, $\rho \sim$0.8, with $p-value <$0.004), but not with $M_{\rm env}$.
	Based on the Pearson $r$ and Spearman $\rho$ coefficients, there is no correlation between number of protostellar components and mass for the multiples subsample in classification scheme 4.
	However, this could be again due to the small sample size. 
	Figure~\ref{fig:numcomp} plots the masses versus number of components.

	\section{Discussion}
	\label{sec:discussion}
	
	\subsection{Physical parameters that influence multiplicity}
	\label{subsec:physvsMultiple}
	In this paper we have used single-dish observations of molecular line emission toward five subregions in the Perseus molecular cloud to derive kinetic gas temperature, volumetric and column densities, and gas masses at protostellar envelope scales.
	These physical parameters have been proposed, through theory, models or previous observations, as key in determining the multiplicity of low-mass protostellar systems. 
	Based on the derived physical parameters and the statistical tests applied to the resulting sample, it is found that kinetic gas temperature, $T_{K}$, is not related to multiplicity, or the number of protostellar sources in a system, in the Perseus molecular cloud.
	Kinetic gas temperature from ammonia, $T_{K,\ce{NH3}}$, does show a relation to gas mass, as well as $L_{\rm bol}$, in NGC1333. But this is more likely due to the method used in estimating the kinetic gas temperature \citep{friesen2017} than an additional correlation.
	Dust temperature, $T_{dust}$, shows correlations to $L_{\rm bol}$ and a weak correlation to n(\ce{H2}).
	The results in this work are consistent with the temperature-multiplicity relation found from APEX observations of \ce{H2CO} and \ce{DCO+} at 217 and 360 GHz \citep{murillo2018}.
	We note that using any of the three temperatures used in this work will lead to practically the same derived \ce{H2} volume density, gas column density, and gas mass. 
	But caution is advised, as implicit dependencies may lead to apparent correlations.

	It is interesting to note that no variation along the outflow direction  (green straight lines in Fig.~\ref{fig:TKmap_N2H}) is seen, especially within the dense regions.
	Outflows have been shown to output sufficient energy and luminosity into the molecular cloud \citep{plunkett2013,dionatos2017}.
	UV-heated gas has been found within outflow cavities \citep{yildiz2015}, and gas temperatures within outflows obtained from \ce{^{13}CO} $J$=10--9/$J$=3--2 $J$=10--9/$J$=6--5 suggest temperatures of $\geq$50 K, assuming a n(\ce{H2}) of 10$^{5}$ -- 10$^{6}$ cm$^{-3}$ \citep{murillo2018}. Hence, even with the $\sim$5 K uncertainty of the $I$(\ce{HCN})/$I$(\ce{HNC}) method, significant temperature variation along the outflow directions should be detected.
	Warm regions ($>$20K) are seen mainly outside of the dense areas traced by \ce{N2H+} and tend to align with outflow directions (e.g., NGC1333: IRAS2, IRAS4, and IRAS7; L448 C), or regions most likely externally irradiated (e.g., L1448 N, IC348).
	
	There is a lack of relation between volumetric \ce{H2} gas density, n(\ce{H2}), and number of protostellar components.
	This is consistent with studies which did not find a correlation between N(\ce{N2H+}) and cloud core density \citep{johnstone2010}.
	The N(\ce{N2H+}) derived in our work match those of \citet{johnstone2010}.
	The same work found a correlation between N(\ce{NH3}) and N(\ce{N2H+}), interpreted as both molecules arising from the same gas, which is also supported by the results of the current work.

	Gas column densities, and consequently gas masses, show strong, positive correlations with multiplicity and number of protostellar components.
	The correlations are consistent whether the multiplicity is within regions of $\sim$5000 AU (18$\arcsec$), or much larger cloud cores.
	In general, cold M$_{gas}$(\ce{N2H+}) is a factor of $\sim$16 higher than warm M$_{gas}$(\ce{HCO+}).
	An explanation regarding the difference in the warm and cold gas masses is that protostars do not heat their envelopes to large extents, hence envelopes mainly contain cold gas.
	Additionally, the angular resolution of the observations is bound to pick up more of the cold gas emission, which would bias the cold to warm gas ratio.
	The former has been shown by models (e.g., \citealt{murillo2022}) and suggested by observations of episodic accretion (e.g., \citealt{hsieh2018,hsieh2019}) and variability (e.g., \citealt{johnstone2022}), with most of the protostellar heating escaping through the outflow cavities. 
	Another possibility is that the gas masses derived in this work are biased due to the constant abundances of \ce{N2H+} and \ce{HCO+} adopted for simplicity in the calculations.
	
	Binaries and singles are found to be drawn from the same parent distribution for M$_{gas}$(\ce{N2H+}), M$_{gas}$(\ce{HCO+}), and $M_{\rm env}$.
	This is true whether the binary subsample consists of only close (separations $\leq$7$\arcsec$) or all systems with two protostellar components.
	The gas and dust mass trends found for single and binary protostellar systems would support the results from analytic models that suggest structures with a few Jeans masses can readily collapse and fragment to form protostars \citep{pon2011}, and earlier studies reporting lower dust masses for close binary systems in comparison to single systems (which would affect disk formation and evolution, \citealt{harris2012}).
	If the dust and gas masses are similar for singles and binaries, another factor must determine whether a single or binary system is formed.

	Our statistical tests indicate that higher order multiple systems (three or more components) are not drawn from the same parent distributions as binaries and singles. 
	Higher order multiples have consistently larger gas and dust masses than binaries and singles.
	It is unclear whether the large gas masses are initially available or accumulated dynamically, and what process stops a cloud core from fragmenting once a few 
	Jeans masses are accumulated.
	\citet{mairs2014} show, based on observations and simulations, that protostars start to form once the cores cross the Jeans unstable threshold and often continue to gain mass from the cloud. 
	However, \citet{mairs2014} did not consider multiplicity due to the inability to follow both mechanical and radiative feedback. 
	Some studies suggest that velocity coherent gas structures in the molecular clouds move material toward protostars \citep{hacar2017,chen2020}, and magnetic fields and turbulence may help in the process of fragmentation (e.g., \citealt{offner2016,lee2019,mathew2021}).
	As mentioned in \citet{lee2019}, these studies do not resolve disks, only turbulent fragmentation is accounted for.
	Disk fragmentation (e.g., \citealt{wurster2019}), and in particular hierarchical fragmentation, could contribute as well.
	However, this is highly dependent on the adopted cooling function and thermodynamic properties of the gas.
	
	While studying these processes is outside the scope of the current work, the higher order multiple systems may provide some insight into this point.
	Within L1448 N, the triple system in component B presents the highest cold gas and dust masses within L1448 N (Fig.~\ref{fig:massColDen}), and has evidence of recent fragmentation \citep{tobin2016N}.
	In contrast, NGC1333 SVS13A has the lowest gas and dust mass in the SVS13 system, but recent studies have reported a small scale accretion flow in NGC1333 SVS13A \citep{diaz2022,hsieh2023}. 
	Toward IRAS03292, \citet{reynolds2023} identify several condensations in dust continuum observations, while \citet{taniguchi2024} recently reported infall from 25000 AU scales onto the cloud core.
	Whether a physical connection exists between envelope scale and molecular cloud scale accretion is still unclear.

	\subsection{Models versus observations}
	\label{subsec:models}
	
	The presence of protostars in cloud cores down to the low end of the sampled range show there was at least a Jeans mass of material during collapse. 
	Analytic studies (e.g., \citealt{pon2011}) reach a similar conclusion, stating that cores with a few Jeans masses can readily fragment if turbulence and magnetic fields are present.
	Prestellar core substructure has been proposed as a mechanism to determine if a core forms a single or multiple protostellar system. 
	Earlier observations did not find such substructure (e.g., \citealt{dunham2016,kirk2017}), however more recent observations report substructure in a starless core (e.g., \citealt{sahu2021}).
	Simulations of turbulent fragmentation models based on observational data (e.g., \citealt{offner2012,dunham2016}) indeed suggest that substructures would not be detectable based on the typical gas densities of the prestellar cores. 
	The results of the current work suggest that amount of mass, rather than density substructures, may be key in fragmentation and formation of multiple protostellar systems.
	While gas mass may not be linked to hierarchy based on models \citep{lee2019}, observations point to a relation between number of protostars and mass.
	If protostellar cloud cores have a mechanism to replenish the gas mass, (e.g., via molecular cloud velocity coherent gas flows; \citealt{hacar2017,chen2020}) or mediated along magnetic field lines \citep{coude2019,doi2020,doi2021}, such replenishment of material could affect the hierarchy of the system through uneven mass distribution.
	Observational evidence shows uneven distribution of dust mass \citep{tobin2010}, and gas mass (e.g., at envelope to disk scales in IRAS16293, \citealt{murillo2022}).
	Speculating even further, cloud cores that get more gas replenishment would lead to the formation of higher order multiple protostellar systems, and have larger (cold) gas mass reservoirs.
	
	The $T_{K}$ maps of Perseus show the inefficiency of low-mass protostars in heating their immediate environment. Hence, heating cannot readily suppress fragmentation at scales beyond a few 100 AU (disk scales), as pointed out in previous work (e.g., \citealt{harsono2011,krumholz2014,murillo2016,offner2022ppvii,mignon-risse2021}).
	Observational evidence of disk fragmentation does exist (e.g., \citealt{tobin2016N}).
	Simulations of low-mass star formation that include outflows (e.g., \citealt{guszejnov2021,mathew2021}), produce lower molecular cloud temperatures since most of the heating from protostars will escape through the outflow cavity, (as shown from observations, e.g., \citealt{yildiz2015}). Such simulations are consistent with the derived kinetic gas temperature maps presented in this work, as well as previous observational work (e.g., \citealt{friesen2017}). However, the current study does not support outflows playing a role in defining stellar masses in protostellar systems \citep{guszejnov2021,mathew2021}.

	Given the inefficiency of heating in suppressing fragmentation, together with similar cloud core masses for single and binary systems, another physical process must be preventing fragmentation. Various models including magnetic fields on $10^3$~AU scales suggest that the magnetic pressure helps stabilize prestellar cores against fragmentation (e.g., the review by \citealt{padoan2014}). 
	Additionally, models of massive star formation considering fragmentation and magnetic fields (e.g., \citealt{mignon-risse2021}) indicate that magnetic pressure dominates over thermal pressure.
	The same is suggested for low-mass star formation if the gas temperature maps presented in the current work along with the magnetic field maps of B1 and NGC1333 \citep{coude2019,doi2020} and inferred magnetic field pressure, are considered.
	However, the gas temperature and magnetic field maps for NGC1333 and B1 need to be directly compared to determine if magnetic pressure dominates over thermal pressure.
	
	\section{Conclusions}
	\label{sec:conclusions}
	We present Nobeyama 45m Radio Observatory OTF maps and APEX single pointing observations of \ce{HNC} $J$=4--3 toward five subregions in the Perseus Molecular Cloud with an angular resolution of $\sim$18$\arcsec$.
	Emission from \ce{HCN} $J$=1--0, \ce{HCO+} $J$=1--0, \ce{HNC} $J$=1--0, \ce{N2H+} $J$=1--0, \ce{^{13}CO} $J$=1--0 and \ce{C^{18}O} $J$=1--0 is detected toward all mapped regions.
	The spatial distribution of each molecular species was compared to the protostellar and starless core population along with outflow directions.
	Gas kinetic temperature maps were derived from the $I$(\ce{HCN})/$I$(\ce{HNC}) $J$=1--0 ratio maps and quantitatively compared to gas and dust temperature maps from literature.
	Molecular hydrogen density was derived from the \ce{HNC} $J$=4--3/$J$=1--0 ratio.
	Using the derived kinetic gas temperature and n(\ce{H2}), column densities and total gas masses were derived from \ce{N2H+} and \ce{HCO+}.
	These quantities provide a physical characterization of protostellar cloud cores at $\sim$5000 AU scales.
	The derived parameters, along with source bolometric luminosity, dust envelope mass, and clustering, were compared with the multiplicity of the protostellar sources in order to determine the factors that influence multiple star formation at molecular cloud scales.
	The following conclusions were drawn from the data.
	\begin{enumerate}
		\item Gas and dust masses are the main factors that define the multiplicity in protostellar systems for the Perseus star forming region at 5000 AU scales. Larger gas and dust masses are needed to produce higher order multiples.
		\item Gas kinetic temperature $T_{K}$, and n(\ce{H2}) do not show a relation to multiplicity, dust or gas mass within the Perseus Molecular Cloud. A weak correlation is found between n(\ce{H2}) and $L_{\rm bol}$, but not between $T_{K}$ and $L_{\rm bol}$.
		\item The continuous relation in gas and dust masses, regardless of how the sample is grouped, suggest a continuum in formation mechanisms rather than distinct formation mechanisms for close and wide multiple protostellar systems.
		\item The results presented here do not support the scenario of outflows setting stellar masses in low-mass star formation.
	\end{enumerate}
	
	The relevance of the gas masses when considering the formation of multiple protostellar systems is demonstrated.
	The physical characterization of cloud cores carried out in this work needs to be repeated at different scales in order to study if the relation between mass and multiplicity is scale dependent. 
	This would be of particular interest when exploring why close binaries show similar masses to single protostellar systems.

	\begin{acknowledgements}
		This paper made use of Nobeyama data, and APEX data.
		The Nobeyama 45-m radio telescope is operated by Nobeyama Radio Observatory, a branch of National Astronomical Observatory of Japan. 
		We are grateful to the APEX staff for support with these observations. Observing time for the APEX data was obtained via Max Planck Institute for Radio Astronomy, Onsala Space Observatory and European Southern Observatory.
		This study is supported by a grant-in-aid from the Ministry of Education, Culture, Sports, Science, and Technology of Japan (20H05645, 20H05845 and 20H05844) and by a pioneering project in RIKEN Evolution of Matter in the Universe (r-EMU).
		The National Radio Astronomy Observatory is a facility of the National Science Foundation operated under cooperative agreement by Associated Universities, Inc.
		N.M.M.~acknowledges support from the RIKEN Special Postdoctoral Researcher Program (Fellowships).
		D.H. is supported by Center for Informatics and Computation in
		Astronomy (CICA) grant and grant number 110J0353I9 from the Ministry
		of Education of Taiwan.
		D.H. acknowledges support from the National Technology and Science
		Council of Taiwan through grant number 111B3005191.
		A.H. has received funding from the European Research Council (ERC) under the European Union’s Horizon 2020 research and innovation programme (Grant agreement No.851435)
		D.J.\ is supported by NRC Canada and by an NSERC Discovery Grant.
		R.M.R.~acknowledges funding from CNES through a postdoctoral fellowship.
		Y.-L.Y. acknowledges support from Grant-in-Aid from the Ministry of Education, Culture, Sports, Science, and Technology of Japan (20H05845, 20H05844, 22K20389), and a pioneering project in RIKEN (Evolution of Matter in the Universe).
	\end{acknowledgements}

	\begin{appendix}
		\section{Observational data}
		\label{ap:obsdata}
\begin{figure*}
			\centering
			\includegraphics[width=0.95\linewidth]{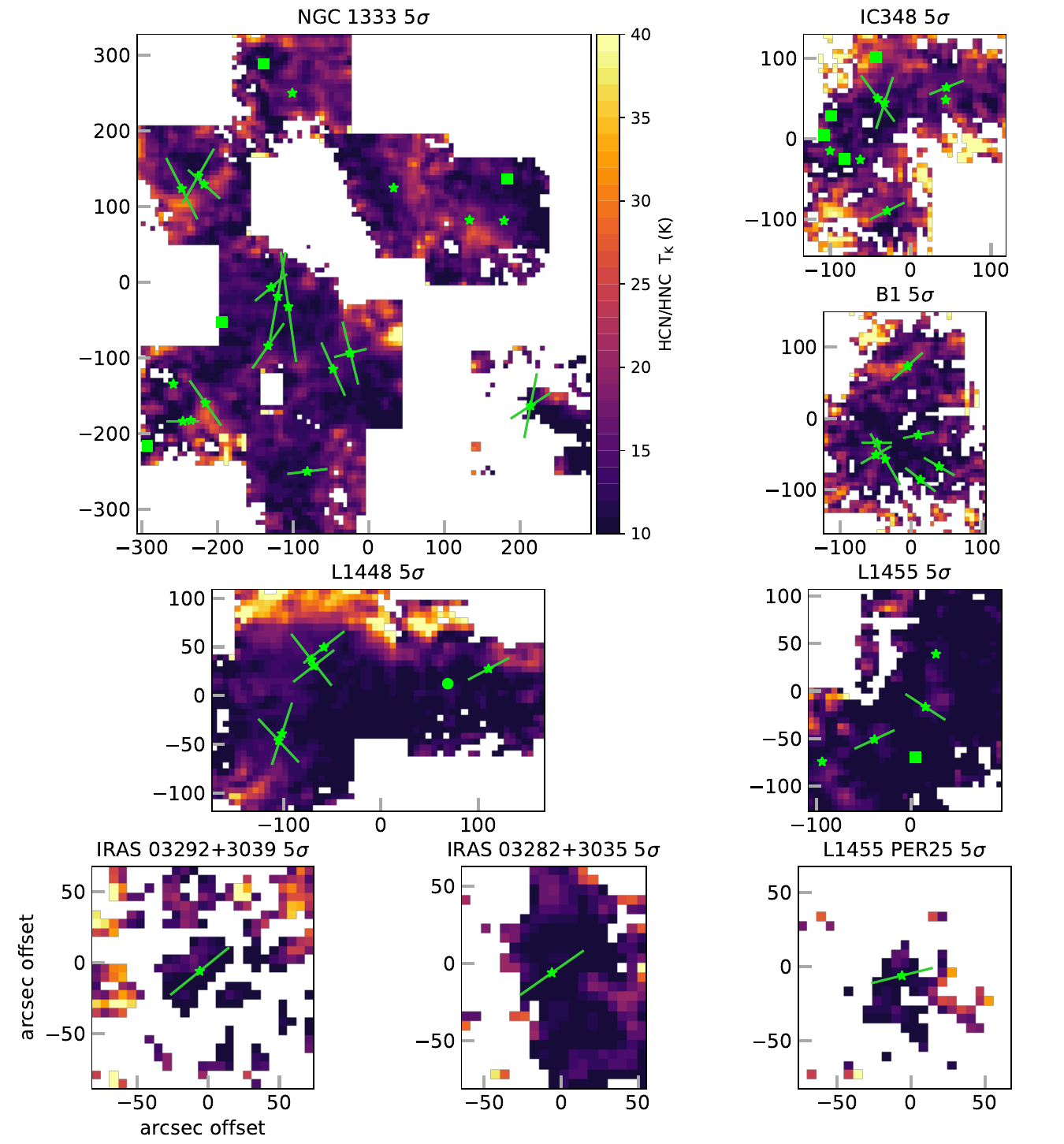}
			\caption{Kinetic temperature maps obtained from the \ce{HCN} 1--0 / \ce{HNC} 1--0 ratio for all observed subregions in Perseus. All maps are shown with the same colorscale range for comparison.  Star symbols mark the locations of protostellar systems (Table~\ref{tab:source}). Square symbols mark the location of starless cores (Table~\ref{tab:starless}). The circle symbol marks the position of L1448 IRS2E, whose nature is debated. Straight lines indicate outflow directions for the protostellar systems included in the MASSES survey \citep{stephens2017}.}
			\label{fig:TKmap}
		\end{figure*}

		The parameters for the protostellar systems and starless cores are listed in Tables~\ref{tab:source} and ~\ref{tab:starless}, respectively.
		Transitions, frequencies, upper energy levels E$_{\rm up}$ and Einstein coefficients A$_{\rm ij}$ for the molecular species presented in this work are listed in Table~\ref{tab:lines}.
		Gas kinetic temperature derived from the \ce{HCN}/\ce{HNC} ratio (see map in Fig.~\ref{fig:TKmap}) are given in Table~\ref{tab:derivedT}.
		The listed values are the mean value within a region of 18$\arcsec$ diameter centered on the positions given in Table~\ref{tab:source}.
		Table~\ref{tab:derivedT} also lists the dust temperature from \citet{zari2016} and \ce{NH3} kinetic temperature from \citet{friesen2017} for the same positions and region size.
		The derived \ce{H2} densities, \ce{N2H+} and \ce{HCO+} column densities and gas masses, together with the dust envelope masses derived from the 850 $\mu$m continuum maps \citep{arce2010} are listed in Table~\ref{tab:derivednNM}.
		The data from these tables is used to plot Figures~\ref{fig:pointplots} and \ref{fig:massColDen}.
		
		\begin{table*} 
			\centering
			\caption{Sample of protostellar systems}
			\begin{tabular}{c c c c c c c}
				\hline \hline
				System & Sources & RA & Dec & Projected  & Clustered?\tablefootmark{a} & $L_{\rm bol}$ (L$_{\odot}$) \\
				 &  &  &  & Separation ($\arcsec$) &  & \\
				\hline
				\multicolumn{7}{c}{Wide multiples}\\
				\hline
				L1448 N\tablefootmark{b}	&	A	&	03:25:36.53	&	30:45:21.35	&	…	&	N	&	8.24	$\pm$	1.29 \\
				&	B	&	03:25:36.34	&	30:45:14.94	&	7.3	&		&	2.15	$\pm$	0.33 \\
				&	C	&	03:25:35.53	&	30:45:34.20	&	16.3	&		&	2.95	$\pm$	0.45 \\
				L1448 C	&	N	&	03:25:38.87	&	30:44:05.40	&	…	&	N	&	0.96	$\pm$	0.16 \\
				&	S	&	03:25:39.14	&	30:43:58.30	&	8.1	&		&	1.69	$\pm$	0.27 \\
				NGC1333 IRAS7\tablefootmark{b}	&	  PER18	&	03:29:11.26	&	31:18:31.08	&	…	&	Y	&	4.77	$\pm$	0.73 \\
				&	  PER21	&	03:29:10.67	&	31:18:20.18	&	13.3	&		&	3.50	$\pm$	0.54 \\
				&	  PER49	&	03:29:12.96	&	31:18:14.31	&	27.5	&		&	0.65	$\pm$	0.10 \\
				NGC1333 IRAS4B\tablefootmark{e}	&	B	&	03:29:12.01	&	31:13:08.10	&	…	&	Y	&	8.74	$\pm$	1.36 \\
				&	B’	&	03:29:12.83	&	31:13:06.90	&	10.6	&		&	0.02	$\pm$	0.002 \\
				NGC1333 SVS13\tablefootmark{b}	&	A	&	03:29:03.75	&	31:16:03.76	&	…	&	Y	&	117.47	$\pm$	18.19 \\
				&	B	&	03:29:03.07	&	31:15:52.02	&	14.9	&		&	10.26	$\pm$	1.57 \\
				&	C	&	03:29:01.96	&	31:15:38.26	&	34.7	&		&	2.26	$\pm$	0.35 \\
				NGC1333 IRAS2	&	A	&	03:28:55.57	&	31:14:37.22	&	…	&	Y	&	47.11	$\pm$	7.22 \\
				&	B	&	03:28:57.35	&	31:14:15.93	&	31.4	&		&	5.89	$\pm$	0.91 \\
				B1-b\tablefootmark{b}	&	S	&	03:33:21.30	&	31:07:27.40	&	…	&	N	&	0.35	$\pm$	0.05 \\
				&	N	&	03:33:21.20	&	31:07:44.20	&	17.4	&		&	0.17	$\pm$	0.03 \\
				&	W	&	03:33:20.30	&	31:07:21.29	&	13.9	&		&	0.12	$\pm$	0.02 \\
				B1 P6+P10	&	  Per6	&	03:33:14.40	&	31:07:10.88	&	…	&	N	&	0.21	$\pm$	0.03 \\
				&	  Per10	&	03:33:16.45	&	31:06:52.49	&	31.9	&		&	0.47	$\pm$	0.07 \\
				IC348MMS	&	  Per11	&	03:43:57.06	&	32:03:04.60	&	…	&	N	&	2.34	$\pm$	0.36 \\
				&	E	&	03:43:57.73	&	32:03:10.10	&	10.2	&		&	0.10	$\pm$	0.03 \\
				IC348SMM2	&	S	&	03:43:51.08	&	32:03:08.32	&	…	&	N	&	1.69	$\pm$	0.27 \\
				&	N	&	03:43:51.00	&	32:03:23.76	&	16.1	&		&	0.96	$\pm$	0.16 \\
				IC348 P32+ED366	&	  Per32	&	03:44:02.40	&	32:02:04.89	&	…	&	N	&	0.17	$\pm$	0.03 \\
				&	  ED366	&	03:43:59.44	&	32:01:53.99	&	36.6	&		&	1.30	$\pm$	0.21 \\
				\hline
				\multicolumn{7}{c}{Close binaries}\\
				\hline
				Per17	&		&	03:27:39.09	&	30:13:03.00	&	0.273	&	N	&	8.49	$\pm$	1.31 \\
				L1455 FIR2	&		&	03:27:38.23	&	30:13:58.80	&	0.346	&	N	&	0.64	$\pm$	0.10 \\
				NGC1333 IRAS4A\tablefootmark{e}	&	A1 \& A2	&	03:29:10.51	&	31:13:31.01	&	1.8	&	Y	&	10.17	$\pm$	1.56 \\
				NGC1333 IRAS1\tablefootmark{b}	&	N \& S	&	03:28:37.00	&	31:13:27.00	&	1.9	&	Y	&	11.00	$\pm$	1.78 \\
				EDJ2009-156	&		&	03:28:51.11	&	31:18:15.41	&	3.192	&	Y	&	0.52	$\pm$	0.08 \\
				IRAS 03282+3035\tablefootmark{b,c}	&		&	03:31:21.00	&	30:45:30.00	&	0.098	&	N	&	1.49	$\pm$	0.23 \\
				IRAS 03292+3039	&		&	03:32:17.95	&	30:49:47.60	&	0.085	&	N	&	1.07	$\pm$	0.27 \\
				B1-a	&		&	03:33:16.66	&	31:07:55.20	&	0.391	&	N	&	1.40	$\pm$	0.23 \\
				
				\hline
				\multicolumn{7}{c}{Single systems}\\
				\hline
				L1455 IRS4	&		&	03:27:43.23	&	30:12:28.80	&	…	&	N	&	1.75	$\pm$	0.27 \\
				L1455 IRS2\tablefootmark{e}	&		&	03:27:47.49	&	30:12:05.32	&	…	&	N	&	2.17	$\pm$	0.35 \\
				L1455 Per25\tablefootmark{b}	&		&	03:26:37.46	&	30:15:28.01	&	…	&	N	&	1.09	$\pm$	0.17 \\
				L1448IRS2	&	 	&	03:25:22.40	&	30:45:12.00	&	…	&	N	&	4.33	$\pm$	0.67 \\
				L1448IRS2E  &	  	&	03:25:25.66	&	30:44:56.70	&	…	&	N	&	0.12	$\pm$	0.02 \\
				NGC1333 IRAS5\tablefootmark{b}	&	  PER52	&	03:28:39.72	&	31:17:31.89	&	…	&	Y	&	0.14	$\pm$	0.02 \\
				NGC1333 IRAS5\tablefootmark{b}  &	  PER63	&	03:28:43.28	&	31:17:32.90	&	…	&	Y	&	1.43	$\pm$	0.22	\\
				RNO 15 FIR	&		&	03:29:04.05	&	31:14:46.61	&	…	&	Y	&	0.29	$\pm$	0.05 \\
				NGC1333 SK1\tablefootmark{b}	&		&	03:29:00.52	&	31:12:00.68	&	…	&	Y	&	0.71	$\pm$	0.11 \\
				NGC1333 IRAS6	&		&	03:29:01.57	&	31:20:20.69	&	…	&	Y	&	12.04	$\pm$	1.86 \\
				NGC1333 IRAS4C\tablefootmark{e}	&		&	03:29:13.52	&	31:13:58.01	&	…	&	Y	&	0.73	$\pm$	0.11 \\
				B1-c	&		&	03:33:17.85	&	31:09:32.00	&	…	&	N	&	5.18	$\pm$	0.80 \\
				HH211	&		&	03:43:56.80	&	32:00:50.21	&	0.3	&	N	&	1.61	$\pm$	0.25 \\
				
				\hline
			\end{tabular}
			\\
			\tablefoot{\tablefoottext{a}{N=No; Y=Yes. Clustered regions have 34 YSO~pc$^{-1}$, while non-clustered regions have 6 YSO~pc$^{-1}$, \citep{plunkett2013}.}
				\tablefoottext{b}{Systems from the sample in \citet{murillo2018}.}
				\tablefoottext{c}{\citet{reynolds2023} raise the possibility that IRAS 03282+3035 is not a close binary given recent observations, but cannot confirm. But given the results of the current work, there is no difference in classifying the system as a single or close binary in terms of physical parameters.}
				\tablefoottext{d}{Possibly unresolved binary \citep{tobin2016}.}
				\tablefoottext{e}{No APEX \ce{HNC} $J$=4--3 observations available. For NGC1333 IRAS4, JCMT \ce{HNC} $J$=4--3 is available from \citet{koumpia2016}.}}
			\label{tab:source}
		\end{table*}
		
		\begin{table*} 
			\centering
			\caption{Sample of Starless cores}
			\begin{tabular}{c c c c c c}
				\hline \hline
				Region & Core & RA\tablefootmark{a} & Dec\tablefootmark{a} & Clustered?\tablefootmark{b} & $L_{\rm bol}$ (L$_{\odot}$)\tablefootmark{a} \\
				\hline
				IC348 & SC16 & 03:44:01.0 & 32:01:54.8 & Y & $<$1.6 \\
				IC348 & SC17 & 03:43:57.9 & 32:04:01.5 & Y & $<$2.0 \\
				IC348 & SC18 & 03:44:03.0 & 32:02:24.3 & Y & $<$0.6 \\
				IC348 & SC21 & 03:44:02.3 & 32:02:48.5 & Y & $<$1.2 \\
				L1455 & SC40 & 03:27:39.9 & 30:12:09.8 & N & $<$0.6 \\
				NGC1333 & SC51\tablefootmark{c} & 03:29:08.8 & 31:15:18.1 & Y & $<$1.5\\
				NGC1333 & SC53 & 03:29:04.5 & 31:20:59.1 & Y & $<$3.7 \\
				NGC1333 & SC59 & 03:29:16.5 & 31:12:34.6 & Y & $<$1.3 \\
				NGC1333 & SC60 & 03:28:39.4 & 31:18:27.1 & Y & $<$1.2 \\
				\hline
			\end{tabular}
			\\
			\tablefoot{\tablefoottext{a}{Source coordinates and bolometric luminosities from \citet{hatchell2007SEDs}.}
				\tablefoottext{b}{N=No; Y=Yes. Clustered regions have 34 YSO~pc$^{-1}$, while non-clustered regions have 6 YSO~pc$^{-1}$, \citep{plunkett2013}.} 
				\tablefoottext{c}{Located at the edge of the small map centered on NGC1333 SVS13.}}
			\label{tab:starless}
		\end{table*}
		
		\begin{table*}
			\centering
			\caption{Molecular species in this work}
			\begin{tabular}{c c c c c}
				\hline \hline
				Molecule & Transition & Frequency & E$_{\rm up}$ & log$_{10}$ A$_{\rm ij}$ \\
				& & GHz & K & \\
				\hline
				\ce{HCN} & J=1--0 F=1--1 & 88.63042 & 4.25 & -4.62 \\
				& J=1--0 F=2--1 & 88.63185 & 4.25 & -5.10 \\
				& J=1--0 F=1--0 & 88.63394 & 4.25 & -5.10 \\
				\ce{HCO+} & 1--0 & 89.18853 & 4.28 & -4.38 \\
				\ce{HNC} & 1--0 & 90.66356 & 4.35 & -4.57 \\
				\ce{N2H+} & J=1-0,F1=1-1,F=0-1 & 93.1716157 & 4.47156 & -4.44039 \\
				& J=1-0,F1=1-1,F=2-1 & 93.1719106 & 4.47158 & -5.24996 \\
				& J=1-0,F1=1-1,F=2-2 & 93.1719106 & 4.47158 & -4.51356 \\
				& J=1-0,F1=1-1,F=1-0 & 93.1720477 & 4.47158 & -4.73371 \\
				& J=1-0,F1=1-1,F=1-1 & 93.1720477 & 4.47158 & -5.36331 \\
				& J=1-0,F1=1-1,F=1-2 & 93.1720477 & 4.47158 & -4.87031 \\
				& J=1-0,F1=2-1,F=2-1 & 93.1734734 & 4.47165 & -4.51355 \\
				& J=1-0,F1=2-1,F=2-2 & 93.1734734 & 4.47165 & -5.24995 \\
				& J=1-0,F1=2-1,F=3-2 & 93.1737699 & 4.47166 & -4.44038 \\
				& J=1-0,F1=2-1,F=1-0 & 93.173964 & 4.47167 & -4.945 \\
				& J=1-0,F1=2-1,F=1-1 & 93.173964 & 4.47167 & -4.6289 \\
				& J=1-0,F1=2-1,F=1-2 & 93.173964 & 4.47167 & -5.8456 \\
				& J=1-0,F1=0-1,F=1-0 & 93.1762595 & 4.47178 & -5.18949 \\
				& J=1-0,F1=0-1,F=1-1 & 93.1762595 & 4.47178 & -5.07339 \\
				& J=1-0,F1=0-1,F=1-2 & 93.1762595 & 4.47178 & -4.67019 \\
				\ce{C^{18}O} & 1--0 & 109.78217 & 5.26 & -7.20 \\
				\ce{^{13}CO} & 1--0 & 110.20135 & 5.29 & -7.19 \\
				\ce{HNC} & 4--3 & 362.63030 & 43.51 & -2.64 \\
				\hline
			\end{tabular}
			\\
			\tablefoot{\tablefoottext{a}{Contains both ortho- and para forms.}}
		\tablebib{All rest frequencies were taken from the Cologne Database for Molecular Spectroscopy (CDMS; \citealt{CDMS_2016}).
			Astronomically observed transitions of \ce{HCN} are listed in \citet{thorwirth2003}.
			The \ce{HCO+} entry is based on \citet{woods1975}.
			The entries for \ce{HNC} are based on \citet{hnc1993}.
			The \ce{^{13}CO} entry is based on \citet{13co2000}. The \ce{C^{18}O} entry is based on \citet{c18o1985}.
			The \ce{N2H+} entry is based on \citet{caselli1995,pagani2009}.}
		\label{tab:lines}
	\end{table*}

		\begin{table*}
			\centering
			\caption{Gas Kinetic and Dust temperatures}
			\begin{tabular}{c c c c c}
				\hline \hline
				Position & $I$(\ce{HCN})/$I$(\ce{HNC}) $J$=1--0\tablefootmark{a} & $T_{K}$\tablefootmark{a}	 & $T_{dust}$\tablefootmark{b} & $T_{K,\ce{NH3}}$\tablefootmark{c} \\
				& & K & K & K \\
				\hline
				L1448 C N & 1.29 $\pm$ 0.13 & 15.0 $\pm$ 1.26 & 20.66 $\pm$ 0.6 & ... \\
				L1448 C S & 1.34 $\pm$ 0.13 & 15.0 $\pm$ 1.31 & 21.42 $\pm$ 0.65 & ... \\
				L1448 N B & 1.3 $\pm$ 0.08 & 15.0 $\pm$ 0.78 & 16.78 $\pm$ 0.43 & ... \\
				L1448 N A & 1.31 $\pm$ 0.07 & 15.0 $\pm$ 0.68 & 17.22 $\pm$ 0.48 & ... \\
				L1448 N C & 1.25 $\pm$ 0.07 & 15.0 $\pm$ 0.65 & 17.03 $\pm$ 0.35 & ... \\
				L1448 IRS2 & 1.21 $\pm$ 0.18 & 15.0 $\pm$ 1.79 & 18.06 $\pm$ 0.34 & ... \\
				L1448 IRS2E & 1.15 $\pm$ 0.16 & 15.0 $\pm$ 1.61 & 12.42 $\pm$ 0.12 & ... \\
				L1455 FIR2 & 1.12 $\pm$ 0.09 & 15.0 $\pm$ 0.92 & 14.61 $\pm$ 0.19 & ... \\
				L1455 Per17 & 1.32 $\pm$ 0.08 & 15.0 $\pm$ 0.82 & 21.19 $\pm$ 0.62 & ... \\
				L1455 IRS4 & 1.26 $\pm$ 0.08 & 15.0 $\pm$ 0.8 & 14.53 $\pm$ 0.24 & ... \\
				L1455 Per25 & 1.25 $\pm$ 0.37 & 15.0 $\pm$ 3.68 & 16.17 $\pm$ 0.31 & ... \\
				L1455 IRS2 & 1.13 $\pm$ 0.07 & 15.0 $\pm$ 0.75 & 14.02 $\pm$ 0.13 & ... \\
				L1455 SC40 & 0.78 $\pm$ 0.07 & 15.0 $\pm$ 0.71 & 11.54 $\pm$ 0.1 & ... \\
				NGC1333 SVS13C & 1.18 $\pm$ 0.11 & 15.0 $\pm$ 1.09 & 18.51 $\pm$ 0.41 & 16.99 $\pm$ 0.21 \\
				NGC1333 SVS13B & 1.2 $\pm$ 0.12 & 15.0 $\pm$ 1.16 & 20.79 $\pm$ 0.53 & 16.76 $\pm$ 0.19 \\
				NGC1333 SVS13A & 1.18 $\pm$ 0.12 & 15.0 $\pm$ 1.23 & 23.03 $\pm$ 0.7 & 16.72 $\pm$ 0.22 \\
				NGC1333 IRAS2A & 1.46 $\pm$ 0.2 & 15.0 $\pm$ 2.0 & 30.8 $\pm$ 1.53 & 17.12 $\pm$ 0.32 \\
				NGC1333 IRAS2B & 1.45 $\pm$ 0.22 & 15.0 $\pm$ 2.16 & 18.59 $\pm$ 1.03 & 14.89 $\pm$ 0.36 \\
				NGC1333 IRAS7 Per18 & 1.18 $\pm$ 0.14 & 15.0 $\pm$ 1.36 & 19.03 $\pm$ 0.3 & 13.94 $\pm$ 0.21 \\
				NGC1333 IRAS7 Per21 & 1.62 $\pm$ 0.24 & 16.21 $\pm$ 2.42 & 18.99 $\pm$ 0.42 & 14.24 $\pm$ 0.34 \\
				NGC1333 IRAS7 Per49 & 1.56 $\pm$ 0.28 & 15.57 $\pm$ 2.78 & 14.74 $\pm$ 0.27 & 12.94 $\pm$ 0.32 \\
				NGC1333 IRAS4A & 2.08 $\pm$ 0.17 & 20.78 $\pm$ 1.71 & 17.0 $\pm$ 0.39 & 15.56 $\pm$ 0.32 \\
				NGC1333 IRAS4C & 1.39 $\pm$ 0.28 & 15.0 $\pm$ 2.85 & 14.67 $\pm$ 0.23 & 12.36 $\pm$ 0.21 \\
				NGC1333 IRAS4B & 1.9 $\pm$ 0.21 & 19.03 $\pm$ 2.1 & 16.29 $\pm$ 0.36 & 13.3 $\pm$ 0.33 \\
				NGC1333 IRAS4B' & 1.33 $\pm$ 0.17 & 15.0 $\pm$ 1.7 & 16.61 $\pm$ 0.43 & 12.46 $\pm$ 0.29 \\
				NGC1333 IRAS1 & 0.87 $\pm$ 0.23 & 15.0 $\pm$ 2.26 & 25.47 $\pm$ 0.9 & 14.39 $\pm$ 0.65 \\
				NGC1333 EDJ156 & 1.36 $\pm$ 0.21 & 15.0 $\pm$ 2.05 & 13.83 $\pm$ 0.14 & 15.2 $\pm$ 0.57 \\
				NGC1333 RNO15FIR & 1.49 $\pm$ 0.15 & 15.0 $\pm$ 1.54 & 14.04 $\pm$ 0.13 & 14.89 $\pm$ 0.27 \\
				NGC1333 IRAS6 & 1.6 $\pm$ 0.19 & 15.97 $\pm$ 1.9 & 19.33 $\pm$ 0.35 & 16.11 $\pm$ 0.23 \\
				NGC1333 SK1 & 1.21 $\pm$ 0.17 & 15.0 $\pm$ 1.75 & 15.45 $\pm$ 0.18 & 12.23 $\pm$ 0.34 \\
				NGC1333 IRAS5 Per63 & 1.82 $\pm$ 0.53 & 18.2 $\pm$ 5.3 & 15.09 $\pm$ 0.14 & 10.18 $\pm$ 0.46 \\
				NGC1333 IRAS5 Per52 & 1.37 $\pm$ 0.21 & 15.0 $\pm$ 2.15 & 13.12 $\pm$ 0.12 & 11.77 $\pm$ 0.18 \\
				NGC1333 SC51 & 1.18 $\pm$ 0.1 & 15.0 $\pm$ 0.95 & 12.09 $\pm$ 0.05 & 12.4 $\pm$ 0.14 \\
				NGC1333 SC53 & 1.11 $\pm$ 0.18 & 15.0 $\pm$ 1.77 & 16.68 $\pm$ 0.12 & 14.64 $\pm$ 0.25 \\
				NGC1333 SC59 & 0.8 $\pm$ 0.18 & 15.0 $\pm$ 1.8 & 12.27 $\pm$ 0.05 & 11.71 $\pm$ 0.15 \\
				NGC1333 SC60 & 1.29 $\pm$ 0.24 & 15.0 $\pm$ 2.36 & 11.28 $\pm$ 0.08 & 10.93 $\pm$ 0.43 \\
				B1 bW & 1.4 $\pm$ 0.19 & 15.0 $\pm$ 1.93 & 12.09 $\pm$ 0.12 & ... \\
				B1 bN & 1.36 $\pm$ 0.16 & 15.0 $\pm$ 1.59 & 11.9 $\pm$ 0.1 & ... \\
				B1 bS & 1.67 $\pm$ 0.22 & 16.7 $\pm$ 2.2 & 12.09 $\pm$ 0.12 & ... \\
				B1 Per6 & 0.99 $\pm$ 0.24 & 15.0 $\pm$ 2.37 & 12.3 $\pm$ 0.07 & ... \\
				B1 Per10 & 1.18 $\pm$ 0.33 & 15.0 $\pm$ 3.34 & 12.89 $\pm$ 0.11 & ... \\
				B1 a & 1.33 $\pm$ 0.19 & 15.0 $\pm$ 1.91 & 12.99 $\pm$ 0.08 & ... \\
				IRAS 03292+3039 & 1.35 $\pm$ 0.4 & 15.0 $\pm$ 3.96 & 13.72 $\pm$ 0.21 & ... \\
				IRAS 03282+3035 & 1.02 $\pm$ 0.14 & 15.0 $\pm$ 1.42 & 16.66 $\pm$ 0.42 & ... \\
				B1 c & 1.96 $\pm$ 0.28 & 19.61 $\pm$ 2.8 & 16.69 $\pm$ 0.37 & ... \\
				IC348 SMM2N & 1.48 $\pm$ 0.22 & 15.0 $\pm$ 2.25 & 15.19 $\pm$ 0.13 & ... \\
				IC348 SMM2S & 1.78 $\pm$ 0.33 & 17.81 $\pm$ 3.29 & 15.84 $\pm$ 0.13 & ... \\
				IC348 MMSE & 1.35 $\pm$ 0.29 & 15.0 $\pm$ 2.86 & 16.54 $\pm$ 0.23 & ... \\
				IC348 Per11 & 1.46 $\pm$ 0.26 & 15.0 $\pm$ 2.65 & 16.83 $\pm$ 0.28 & ... \\
				IC348 Per32 & 1.56 $\pm$ 0.24 & 15.65 $\pm$ 2.43 & 15.8 $\pm$ 0.11 & ... \\
				IC348 ED366 & 1.32 $\pm$ 0.35 & 15.0 $\pm$ 3.47 & 17.93 $\pm$ 0.17 & ... \\
				IC348 HH211 & 1.71 $\pm$ 0.34 & 17.09 $\pm$ 3.4 & 17.28 $\pm$ 0.35 & ... \\
				IC348 SC16 & 1.25 $\pm$ 0.22 & 15.0 $\pm$ 2.23 & 16.31 $\pm$ 0.13 & ... \\
				IC348 SC17 & 2.58 $\pm$ 0.69 & 25.79 $\pm$ 6.86 & 15.83 $\pm$ 0.09 & ... \\
				IC348 SC18 & 1.76 $\pm$ 0.34 & 17.64 $\pm$ 3.39 & 15.23 $\pm$ 0.08 & ... \\
				IC348 SC21 & 1.47 $\pm$ 0.37 & 15.0 $\pm$ 3.72 & 15.07 $\pm$ 0.05 & ... \\
				\hline
			\end{tabular}
			\\
			\tablefoot{
				\tablefoottext{a}{For ratios $<$1.5, the gas kinetic temperature is arbitrarily set to 15 K. See Section~\ref{subsec:gasT} for details.}
				\tablefoottext{b}{Dust temperature obtained from the \textit{Herschel} and \textit{Planck} maps from \cite{zari2016}.}
				\tablefoottext{c}{Gas kinetic temperature obtained from fitting \ce{NH3} maps from \cite{friesen2017}.}}
			\label{tab:derivedT}
		\end{table*}
		
		\begin{table*}
			\centering
			\caption{Densities and Masses}
			\begin{tabular}{c c c c c c c c}
				\hline \hline
				Position & \ce{HNC}  & $n({\ce{H2}})$\tablefootmark{a} & $N({\ce{N2H+}})$\tablefootmark{a} & $N({\ce{HCO+}})$\tablefootmark{a} & $M_{gas}(\ce{N2H+})$\tablefootmark{a} & $M_{gas}(\ce{HCO+})$\tablefootmark{a} & $M_{env}$ \\
				& $J$=4--3/$J$=1--0 & 10$^{6}$ cm$^{-3}$ & 10$^{13}$ cm$^{-2}$ & 10$^{12}$ cm$^{-2}$ & 10$^{-9}$ M$_{\odot}$ & 10$^{-10}$ M$_{\odot}$ & M$_{\odot}$ \\
				\hline
				L1448 C N & 0.9 $\pm$ 0.04 & 5.66 $\pm$ 4.02 & 2.37 $\pm$ 0.60 & 1.53 $\pm$ 0.18 & 10.90 $\pm$ 2.75 & 6.99 $\pm$ 0.82 & 3.11 \\
				L1448 C S & 0.82 $\pm$ 0.03 & 5.16 $\pm$ 3.08 & 2.28 $\pm$ 0.21 & 1.47 $\pm$ 0.14 & 10.40 $\pm$ 0.95 & 6.74 $\pm$ 0.65 & 2.54 \\
				L1448 N B & 0.21 $\pm$ 0.01 & 1.41 $\pm$ 0.75 & 3.47 $\pm$ 0.17 & 2.17 $\pm$ 0.21 & 15.90 $\pm$ 0.78 & 9.95 $\pm$ 0.97 & 7.33 \\
				L1448 N A & 0.23 $\pm$ 0.01 & 1.24 $\pm$ 0.64 & 3.10 $\pm$ 0.13 & 1.92 $\pm$ 0.19 & 14.20 $\pm$ 0.60 & 8.80 $\pm$ 0.85 & 4.52 \\
				L1448 N C & 0.2 $\pm$ 0.01 & 1.03 $\pm$ 0.62 & 2.53 $\pm$ 0.08 & 1.53 $\pm$ 0.15 & 11.60 $\pm$ 0.37 & 7.00 $\pm$ 0.71 & 4.05 \\
				L1448 IRS2 & 0.16 $\pm$ 0.01 & 0.82 $\pm$ 0.14 & 1.08 $\pm$ 0.02 & 0.66 $\pm$ 0.02 & 4.96 $\pm$ 0.07 & 3.01 $\pm$ 0.11 & 1.20 \\
				L1448 IRS2E & 0.18 $\pm$ 0.02 & 0.80 $\pm$ 0.13 & 1.50 $\pm$ 0.01 & 0.91 $\pm$ 0.03 & 6.87 $\pm$ 0.05 & 4.17 $\pm$ 0.12 & 1.46 \\
				L1455 FIR2 & 0.47 $\pm$ 0.03 & 2.73 $\pm$ 1.72 & 0.55 $\pm$ 0.05 & 0.35 $\pm$ 0.05 & 2.74 $\pm$ 0.27 & 1.75 $\pm$ 0.23 & 0.42 \\
				L1455 Per17 & 0.33 $\pm$ 0.01 & 1.52 $\pm$ 0.84 & 1.09 $\pm$ 0.08 & 0.68 $\pm$ 0.08 & 5.36 $\pm$ 0.37 & 3.37 $\pm$ 0.39 & 0.45 \\
				L1455 IRS4 & 0.24 $\pm$ 0.01 & 1.21 $\pm$ 0.73 & 0.83 $\pm$ 0.05 & 0.51 $\pm$ 0.06 & 4.11 $\pm$ 0.24 & 2.53 $\pm$ 0.31 & 0.56 \\
				L1455 Per25 & 0.34 $\pm$ 0.06 & 1.68 $\pm$ 1.00 & 0.52 $\pm$ 0.05 & 0.33 $\pm$ 0.05 & 2.56 $\pm$ 0.23 & 1.62 $\pm$ 0.25 & 0.34 \\
				L1455 IRS2\tablefootmark{b} & ... & 1.79 $\pm$ 0.00 & 0.85 $\pm$ 0.01 & 0.55 $\pm$ 0.01 & 4.19 $\pm$ 0.04 & 2.71 $\pm$ 0.04 & 0.33 \\
				L1455 SC40\tablefootmark{b} & ... & 1.79 $\pm$ 0.00 & 0.95 $\pm$ 0.01 & 0.61 $\pm$ 0.01 & 4.68 $\pm$ 0.04 & 3.02 $\pm$ 0.04 & 0.56 \\
				NGC1333 SVS13C & 0.17 $\pm$ 0.01 & 0.98 $\pm$ 0.59 & 3.29 $\pm$ 0.09 & 1.98 $\pm$ 0.20 & 16.20 $\pm$ 0.42 & 9.79 $\pm$ 0.96 & 2.73 \\
				NGC1333 SVS13B & 0.31 $\pm$ 0.01 & 1.76 $\pm$ 1.03 & 2.34 $\pm$ 0.17 & 1.48 $\pm$ 0.17 & 11.60 $\pm$ 0.81 & 7.32 $\pm$ 0.85 & 3.41 \\
				NGC1333 SVS13A & 0.49 $\pm$ 0.01 & 2.74 $\pm$ 1.77 & 1.90 $\pm$ 0.18 & 1.22 $\pm$ 0.15 & 9.38 $\pm$ 0.89 & 6.00 $\pm$ 0.75 & 1.42 \\
				NGC1333 IRAS2A & 0.82 $\pm$ 0.03 & 5.23 $\pm$ 3.20 & 1.81 $\pm$ 0.20 & 1.17 $\pm$ 0.14 & 8.95 $\pm$ 0.96 & 5.77 $\pm$ 0.70 & 1.29 \\
				NGC1333 IRAS2B & 0.28 $\pm$ 0.02 & 1.24 $\pm$ 0.69 & 0.89 $\pm$ 0.07 & 0.55 $\pm$ 0.07 & 4.40 $\pm$ 0.32 & 2.72 $\pm$ 0.37 & 0.41 \\
				NGC1333 IRAS7 Per18 & 0.23 $\pm$ 0.01 & 1.12 $\pm$ 0.61 & 1.79 $\pm$ 0.06 & 1.10 $\pm$ 0.10 & 8.82 $\pm$ 0.32 & 5.41 $\pm$ 0.52 & 1.40 \\
				NGC1333 IRAS7 Per21 & 0.38 $\pm$ 0.02 & 1.55 $\pm$ 0.84 & 1.39 $\pm$ 0.12 & 0.89 $\pm$ 0.13 & 6.88 $\pm$ 0.59 & 4.37 $\pm$ 0.62 & 1.57 \\
				NGC1333 IRAS7 Per49 & 0.13 $\pm$ 0.02 & 0.73 $\pm$ 0.46 & 0.75 $\pm$ 0.03 & 0.44 $\pm$ 0.05 & 3.70 $\pm$ 0.15 & 2.17 $\pm$ 0.24 & 1.08 \\
				NGC1333 IRAS4A & 0.14 $\pm$ 0.01 & 0.56 $\pm$ 0.40 & 1.55 $\pm$ 0.09 & 0.89 $\pm$ 0.12 & 7.63 $\pm$ 0.43 & 4.37 $\pm$ 0.57 & 9.17 \\
				NGC1333 IRAS4C & 0.15 $\pm$ 0.01 & 1.06 $\pm$ 0.46 & 1.38 $\pm$ 0.04 & 0.85 $\pm$ 0.07 & 6.81 $\pm$ 0.21 & 4.20 $\pm$ 0.35 & 1.43 \\
				NGC1333 IRAS4B & 0.15 $\pm$ 0.01 & 0.83 $\pm$ 0.53 & 1.30 $\pm$ 0.09 & 0.79 $\pm$ 0.13 & 6.41 $\pm$ 0.44 & 3.89 $\pm$ 0.64 & 4.17 \\
				NGC1333 IRAS4B' & 0.1 $\pm$ 0.01 & 0.60 $\pm$ 0.35 & 1.36 $\pm$ 0.07 & 0.78 $\pm$ 0.05 & 6.73 $\pm$ 0.35 & 3.85 $\pm$ 0.26 & 25.69 \\
				NGC1333 IRAS1 & 0.53 $\pm$ 0.03 & 2.21 $\pm$ 1.10 & 0.61 $\pm$ 0.06 & 0.39 $\pm$ 0.05 & 3.00 $\pm$ 0.28 & 1.92 $\pm$ 0.25 & 0.44 \\
				NGC1333 EDJ156 & 0.11 $\pm$ 0.02 & 0.66 $\pm$ 0.39 & 0.55 $\pm$ 0.02 & 0.32 $\pm$ 0.03 & 2.69 $\pm$ 0.09 & 1.56 $\pm$ 0.14 & 0.17 \\
				NGC1333 RNO15FIR & 0.3 $\pm$ 0.01 & 1.63 $\pm$ 1.04 & 1.75 $\pm$ 0.15 & 1.09 $\pm$ 0.16 & 8.62 $\pm$ 0.74 & 5.39 $\pm$ 0.77 & 1.77 \\
				NGC1333 IRAS6 & 0.39 $\pm$ 0.01 & 1.59 $\pm$ 0.81 & 1.75 $\pm$ 0.12 & 1.11 $\pm$ 0.13 & 8.62 $\pm$ 0.59 & 5.49 $\pm$ 0.64 & 1.00 \\
				NGC1333 SK1 & 0.12 $\pm$ 0.01 & 0.80 $\pm$ 0.44 & 1.58 $\pm$ 0.04 & 0.94 $\pm$ 0.08 & 7.80 $\pm$ 0.21 & 4.64 $\pm$ 0.37 & 0.37 \\
				NGC1333 IRAS5 Per63 & 0.15 $\pm$ 0.03 & 0.85 $\pm$ 0.55 & 0.27 $\pm$ 0.02 & 0.17 $\pm$ 0.03 & 1.35 $\pm$ 0.12 & 0.83 $\pm$ 0.16 & 0.11 \\
				NGC1333 IRAS5 Per52 & 0.24 $\pm$ 0.02 & 1.14 $\pm$ 0.66 & 1.10 $\pm$ 0.06 & 0.68 $\pm$ 0.08 & 5.44 $\pm$ 0.29 & 3.36 $\pm$ 0.41 & 1.00 \\
				NGC1333 SC51\tablefootmark{b} & ... & 1.44 $\pm$ 0.00 & 2.36 $\pm$ 0.02 & 1.50 $\pm$ 0.02 & 11.70 $\pm$ 0.08 & 7.42 $\pm$ 0.08 & 1.22 \\
				NGC1333 SC53\tablefootmark{b} & ... & 1.44 $\pm$ 0.00 & 1.02 $\pm$ 0.01 & 0.65 $\pm$ 0.01 & 5.05 $\pm$ 0.03 & 3.22 $\pm$ 0.05 & 0.81 \\
				NGC1333 SC59\tablefootmark{b} & ... & 1.44 $\pm$ 0.00 & 1.02 $\pm$ 0.02 & 0.65 $\pm$ 0.02 & 5.03 $\pm$ 0.10 & 3.19 $\pm$ 0.10 & 0.60 \\
				NGC1333 SC60\tablefootmark{b} & ... & 1.44 $\pm$ 0.00 & 1.30 $\pm$ 0.03 & 0.83 $\pm$ 0.03 & 6.41 $\pm$ 0.13 & 4.08 $\pm$ 0.14 & 0.62 \\
				B1 bW & 0.16 $\pm$ 0.02 & 1.01 $\pm$ 0.56 & 1.85 $\pm$ 0.07 & 1.13 $\pm$ 0.12 & 9.23 $\pm$ 0.36 & 5.63 $\pm$ 0.60 & 4.48 \\
				B1 bN & 0.13 $\pm$ 0.02 & 0.78 $\pm$ 0.47 & 1.85 $\pm$ 0.06 & 1.09 $\pm$ 0.10 & 9.23 $\pm$ 0.29 & 5.43 $\pm$ 0.49 & 6.12 \\
				B1 bS & 0.18 $\pm$ 0.02 & 0.97 $\pm$ 0.58 & 1.53 $\pm$ 0.09 & 0.94 $\pm$ 0.13 & 7.65 $\pm$ 0.44 & 4.69 $\pm$ 0.66 & 5.08 \\
				B1 Per6 & 0.32 $\pm$ 0.03 & 1.68 $\pm$ 1.00 & 1.83 $\pm$ 0.14 & 1.15 $\pm$ 0.15 & 9.13 $\pm$ 0.71 & 5.76 $\pm$ 0.76 & 1.89 \\
				B1 Per10 & 0.31 $\pm$ 0.04 & 1.63 $\pm$ 0.95 & 1.66 $\pm$ 0.12 & 1.05 $\pm$ 0.14 & 8.30 $\pm$ 0.58 & 5.25 $\pm$ 0.68 & 1.77 \\
				B1 a & 0.15 $\pm$ 0.01 & 0.95 $\pm$ 0.44 & 1.34 $\pm$ 0.04 & 0.82 $\pm$ 0.08 & 6.70 $\pm$ 0.22 & 4.09 $\pm$ 0.38 & 0.55 \\
				B1 IRAS 03292+3039 & 0.39 $\pm$ 0.02 & 2.32 $\pm$ 1.35 & 0.88 $\pm$ 0.10 & 0.56 $\pm$ 0.09 & 4.38 $\pm$ 0.52 & 2.81 $\pm$ 0.47 & 3.58 \\
				B1 IRAS 03282+3035 & 0.29 $\pm$ 0.02 & 1.51 $\pm$ 0.97 & 1.26 $\pm$ 0.08 & 0.79 $\pm$ 0.10 & 6.28 $\pm$ 0.38 & 3.93 $\pm$ 0.49 & 1.33 \\
				B1 c & 0.55 $\pm$ 0.02 & 1.99 $\pm$ 1.23 & 1.75 $\pm$ 0.24 & 1.15 $\pm$ 0.22 & 8.75 $\pm$ 1.23 & 5.74 $\pm$ 1.12 & 2.10 \\
				IC348 SMM2N & 0.27 $\pm$ 0.01 & 1.12 $\pm$ 0.56 & 0.80 $\pm$ 0.04 & 0.49 $\pm$ 0.05 & 3.82 $\pm$ 0.19 & 2.36 $\pm$ 0.25 & 0.78 \\
				IC348 SMM2S & 0.23 $\pm$ 0.01 & 0.93 $\pm$ 0.63 & 0.45 $\pm$ 0.03 & 0.27 $\pm$ 0.05 & 2.16 $\pm$ 0.16 & 1.32 $\pm$ 0.23 & 0.55 \\
				IC348 MMSE & 0.29 $\pm$ 0.01 & 1.63 $\pm$ 1.05 & 0.71 $\pm$ 0.07 & 0.45 $\pm$ 0.07 & 3.43 $\pm$ 0.34 & 2.15 $\pm$ 0.35 & 4.75 \\
				IC348 Per11 & 0.24 $\pm$ 0.01 & 1.20 $\pm$ 0.68 & 0.67 $\pm$ 0.05 & 0.41 $\pm$ 0.06 & 3.21 $\pm$ 0.22 & 1.99 $\pm$ 0.27 & 1.63 \\
				IC348 Per32 & 0.14 $\pm$ 0.01 & 0.74 $\pm$ 0.47 & 0.65 $\pm$ 0.02 & 0.38 $\pm$ 0.04 & 3.10 $\pm$ 0.12 & 1.83 $\pm$ 0.21 & 1.06 \\
				IC348 ED366 & 0.04 $\pm$ 0.01 & 0.35 $\pm$ 0.24 & 0.55 $\pm$ 0.12 & 0.25 $\pm$ 0.03 & 2.62 $\pm$ 0.56 & 1.19 $\pm$ 0.14 & 0.27 \\
				IC348 HH211 & 0.26 $\pm$ 0.02 & 1.20 $\pm$ 0.79 & 1.18 $\pm$ 0.10 & 0.73 $\pm$ 0.13 & 5.67 $\pm$ 0.47 & 3.52 $\pm$ 0.62 & 2.51 \\
				IC348 SC16\tablefootmark{b} & ... & 1.02 $\pm$ 0.00 & 0.89 $\pm$ 0.01 & 0.55 $\pm$ 0.01 & 4.27 $\pm$ 0.05 & 2.64 $\pm$ 0.07 & 0.51 \\
				IC348 SC17\tablefootmark{b} & ... & 1.02 $\pm$ 0.00 & 0.58 $\pm$ 0.04 & 0.38 $\pm$ 0.04 & 2.77 $\pm$ 0.17 & 1.83 $\pm$ 0.19 & 0.35 \\
				IC348 SC18\tablefootmark{b} & ... & 1.02 $\pm$ 0.00 & 0.32 $\pm$ 0.01 & 0.20 $\pm$ 0.01 & 1.53 $\pm$ 0.04 & 0.96 $\pm$ 0.05 & 0.59 \\
				IC348 SC21\tablefootmark{b} & ... & 1.02 $\pm$ 0.00 & 0.47 $\pm$ 0.01 & 0.29 $\pm$ 0.01 & 2.25 $\pm$ 0.03 & 1.40 $\pm$ 0.05 & 0.43 \\
				\hline
			\end{tabular}
			\\
			\tablefoot{\tablefoottext{a}{Column densities, gas masses, and envelope dust masses calculated for circular regions of 18$\arcsec$ in diameter, approximately 5000 AU, centered on the sources listed in Tables~\ref{tab:source} and \ref{tab:starless}.}
			\tablefoottext{b}{Sources without \ce{HNC} $J$=4--3 data, adopted n(\ce{H2}) is an average of the corresponding region, which could be an upper or lower limit.}}
			\label{tab:derivednNM}
		\end{table*}

		\section{Intensity integrated maps}
		\label{ap:moment}
\begin{figure*}
			\centering
			\includegraphics[width=0.9\linewidth]{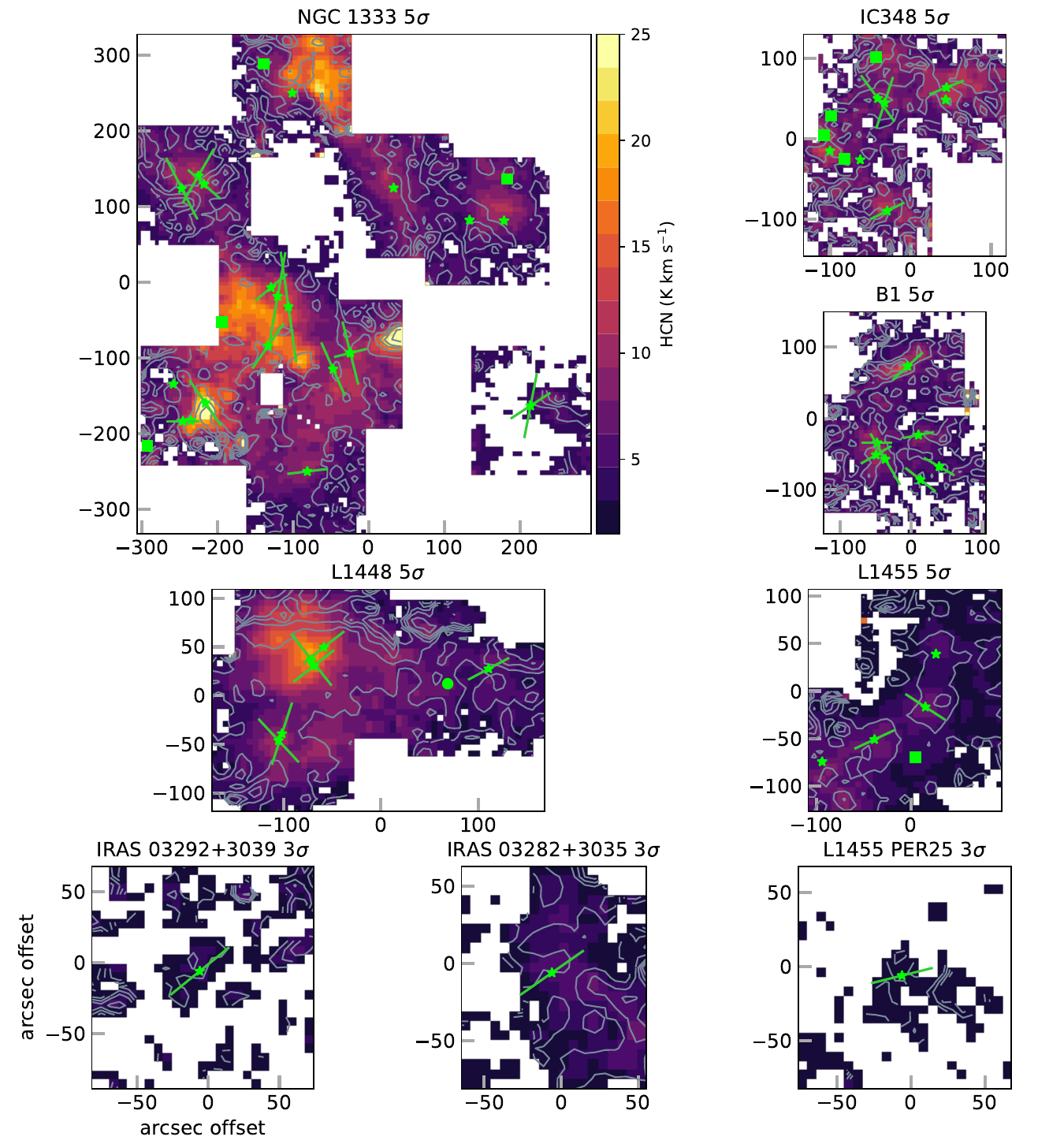}
			\caption{Integrated intensity maps of hydrogen cyanide \ce{HCN} 1--0 toward five subregions in Perseus: NGC1333, L1448, L1455, B1, and IC348. All maps are shown with the same colorscale range for comparison. Maps are clipped to the signal-to-noise ratio indicated next to the panel title (i.e., 5$\sigma$ or 3$\sigma$). Gas kinetic temperature map is overlaid in contours in steps of 10., 15., 20., 25., 30., 40. K. Star symbols mark the locations of protostellar systems (Table~\ref{tab:source}). Square symbols mark the location of starless cores (Table~\ref{tab:starless}). The circle symbol marks the position of L1448 IRS2E, whose nature is debated. Straight lines indicate outflow directions for the protostellar systems included in the MASSES survey \citep{stephens2017}.}
			\label{fig:HCNmom0}
		\end{figure*}

\begin{figure*}
			\centering
			\includegraphics[width=0.9\linewidth]{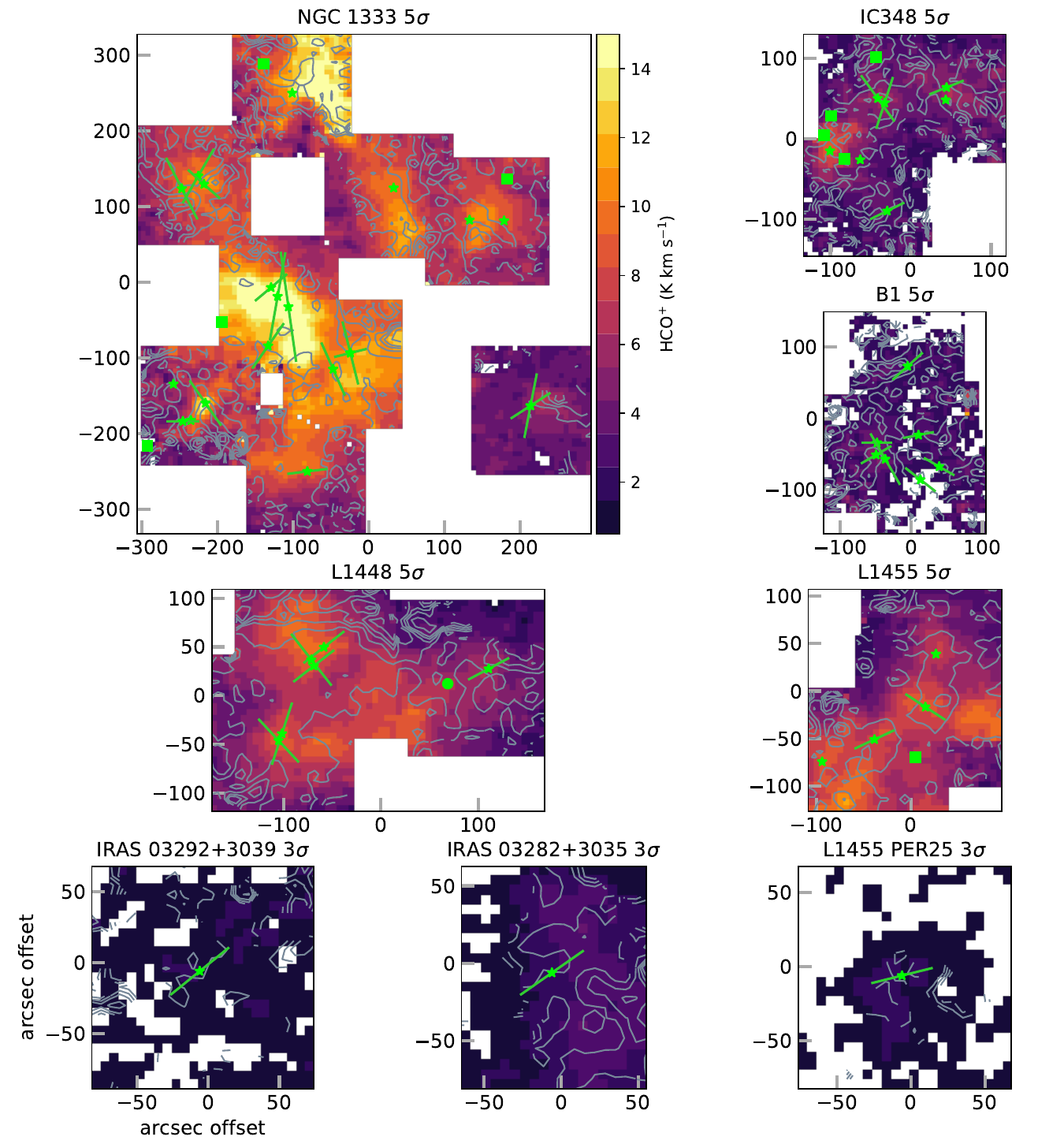}
			\caption{Integrated intensity maps of formyl cation \ce{HCO+} 1--0 toward five subregions in Perseus: NGC1333, L1448, L1455, B1, and IC348. All maps are shown with the same colorscale range for comparison. Maps are clipped to the signal-to-noise ratio indicated next to the panel title (i.e., 5$\sigma$ or 3$\sigma$). Gas kinetic temperature map is overlaid in contours in steps of 10., 15., 20., 25., 30., 40. K. Star symbols mark the locations of protostellar systems (Table~\ref{tab:source}). Square symbols mark the location of starless cores (Table~\ref{tab:starless}). The circle symbol marks the position of L1448 IRS2E, whose nature is debated. Straight lines indicate outflow directions for the protostellar systems included in the MASSES survey \citep{stephens2017}.}
			\label{fig:HCOmom0}
		\end{figure*}

\begin{figure*}
			\centering
			\includegraphics[width=0.9\linewidth]{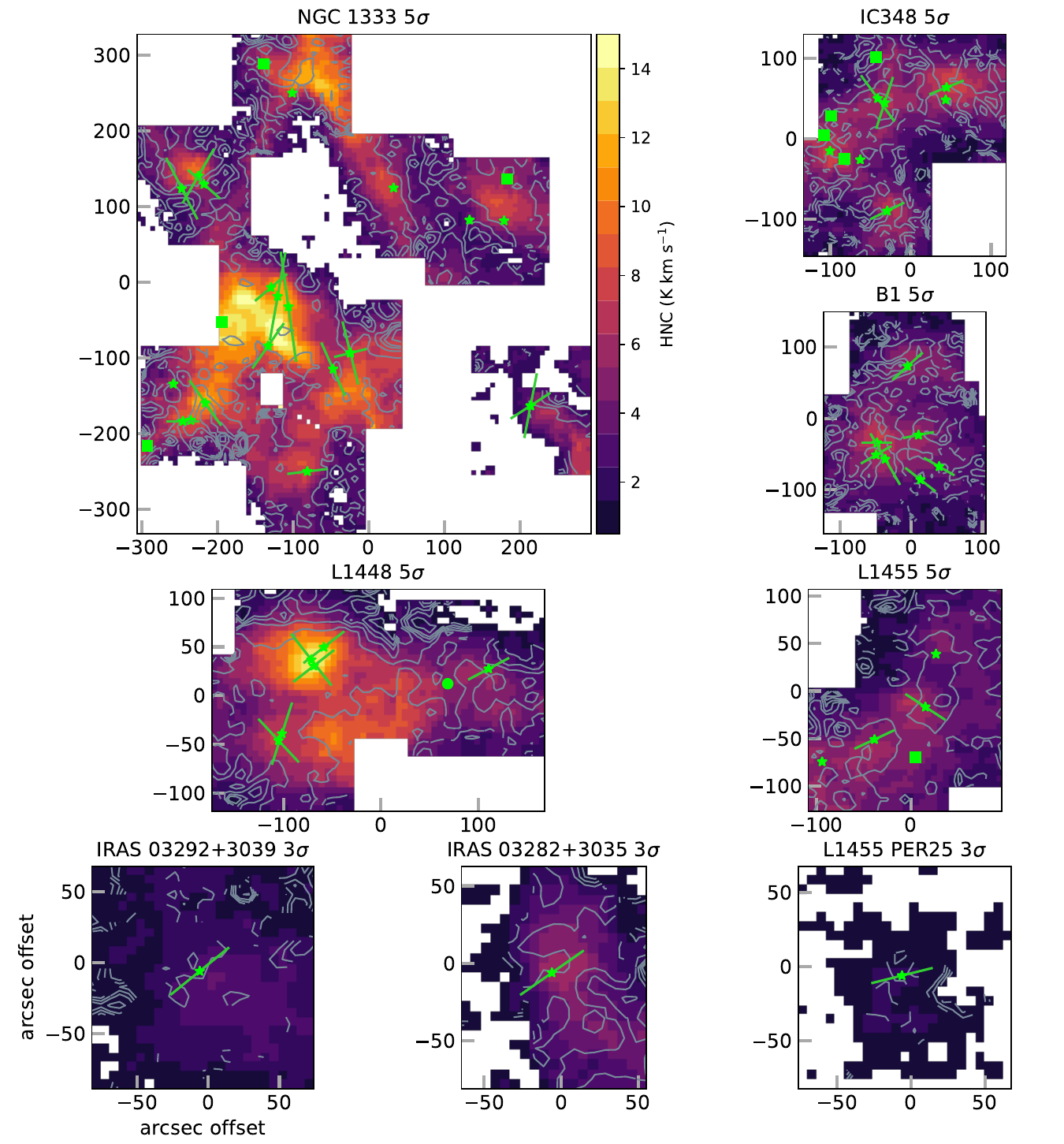}
			\caption{Integrated intensity maps of hydrogen isocyanide \ce{HNC} 1--0 toward five subregions in Perseus: NGC1333, L1448, L1455, B1, and IC348. All maps are shown with the same colorscale range for comparison. Maps are clipped to the signal-to-noise ratio indicated next to the panel title (i.e., 5$\sigma$ or 3$\sigma$). Gas kinetic temperature map is overlaid in contours in steps of 10., 15., 20., 25., 30., 40. K. Star symbols mark the locations of protostellar systems (Table~\ref{tab:source}). Square symbols mark the location of starless cores (Table~\ref{tab:starless}). The circle symbol marks the position of L1448 IRS2E, whose nature is debated. Straight lines indicate outflow directions for the protostellar systems included in the MASSES survey \citep{stephens2017}.}
			\label{fig:HNCmom0}
		\end{figure*}

\begin{figure*}
			\centering
			\includegraphics[width=0.9\linewidth]{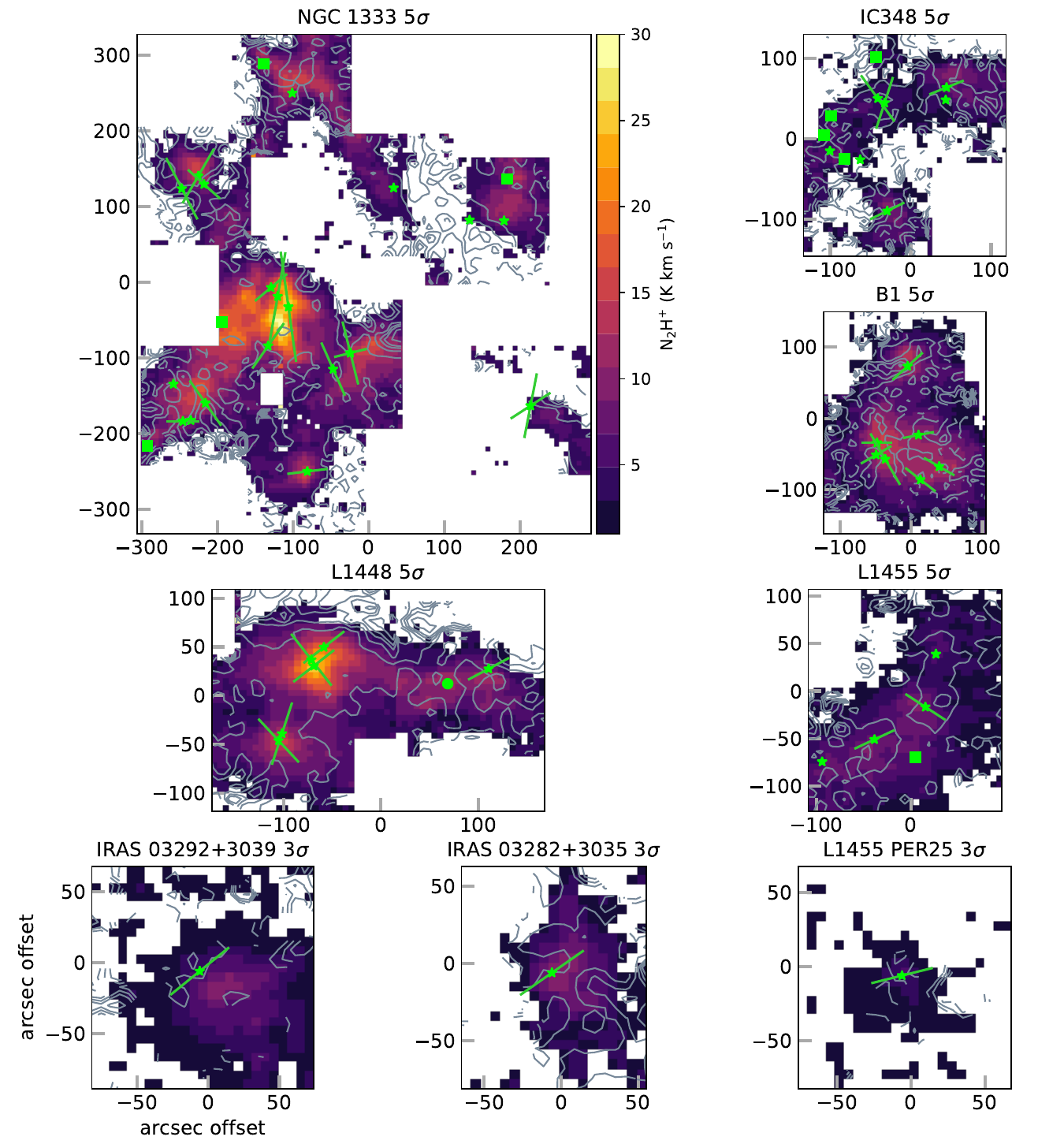}
			\caption{Integrated intensity maps of diazenylium \ce{N2H+} 1--0 toward five subregions in Perseus: NGC1333, L1448, L1455, B1, and IC348. All maps are shown with the same colorscale range for comparison. Maps are clipped to the signal-to-noise ratio indicated next to the panel title (i.e., 5$\sigma$ or 3$\sigma$). Gas kinetic temperature map is overlaid in contours in steps of 10., 15., 20., 25., 30., 40. K. Star symbols mark the locations of protostellar systems (Table~\ref{tab:source}). Square symbols mark the location of starless cores (Table~\ref{tab:starless}). The circle symbol marks the position of L1448 IRS2E, whose nature is debated. Straight lines indicate outflow directions for the protostellar systems included in the MASSES survey \citep{stephens2017}.}
			\label{fig:N2Hmom0}
		\end{figure*}

\begin{figure*}
			\centering
			\includegraphics[width=0.9\linewidth]{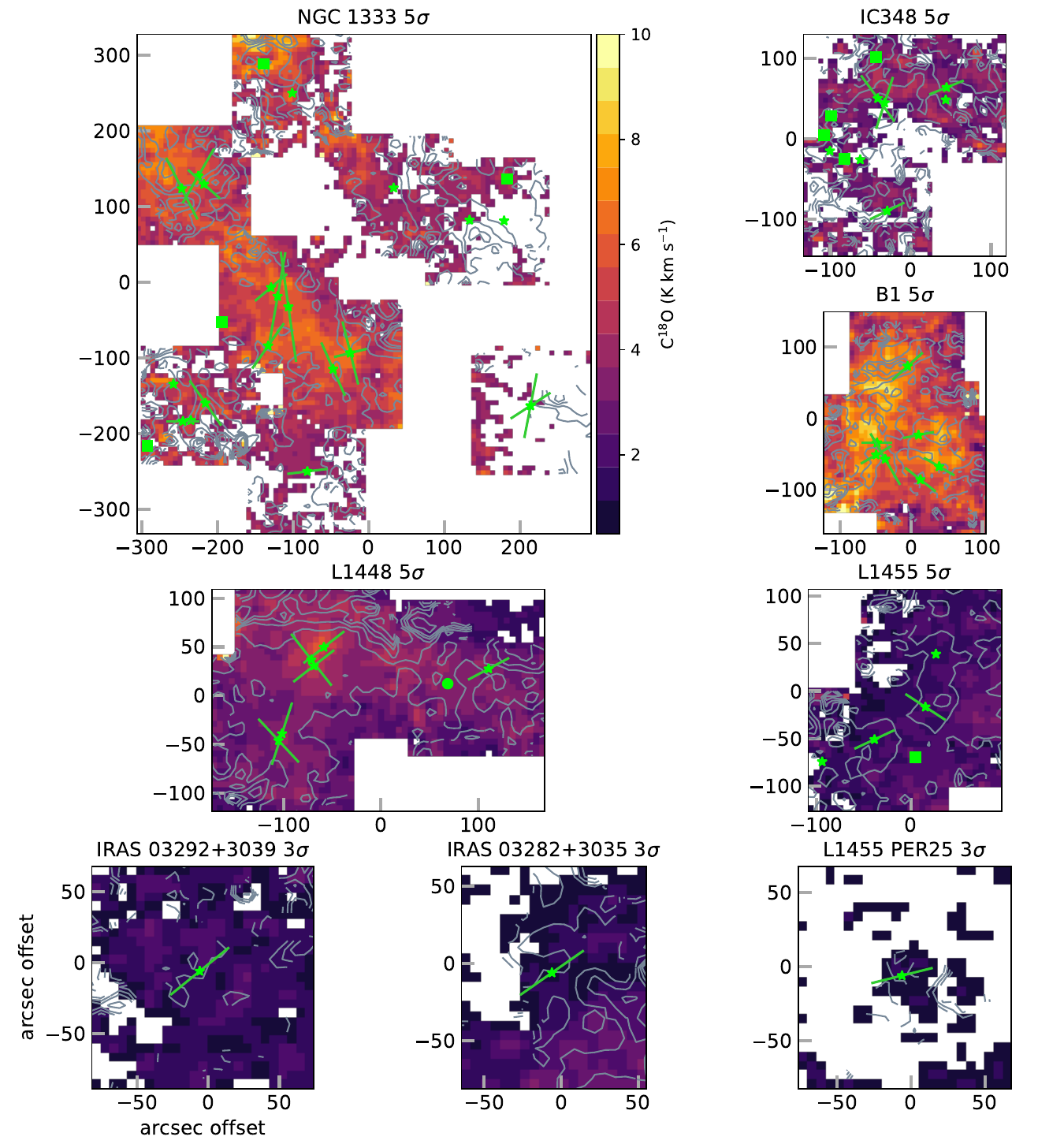}
			\caption{Integrated intensity maps of carbon monoxide \ce{C^{18}O} 1--0 toward five subregions in Perseus: NGC1333, L1448, L1455, B1, and IC348. All maps are shown with the same colorscale range for comparison. Maps are clipped to the signal-to-noise ratio indicated next to the panel title (i.e., 5$\sigma$ or 3$\sigma$). Gas kinetic temperature map is overlaid in contours in steps of 10., 15., 20., 25., 30., 40. K. Star symbols mark the locations of protostellar systems (Table~\ref{tab:source}). Square symbols mark the location of starless cores (Table~\ref{tab:starless}). The circle symbol marks the position of L1448 IRS2E, whose nature is debated. Straight lines indicate outflow directions for the protostellar systems included in the MASSES survey \citep{stephens2017}.}
			\label{fig:C18Omom0}
		\end{figure*}

\begin{figure*}
			\centering
			\includegraphics[width=0.9\linewidth]{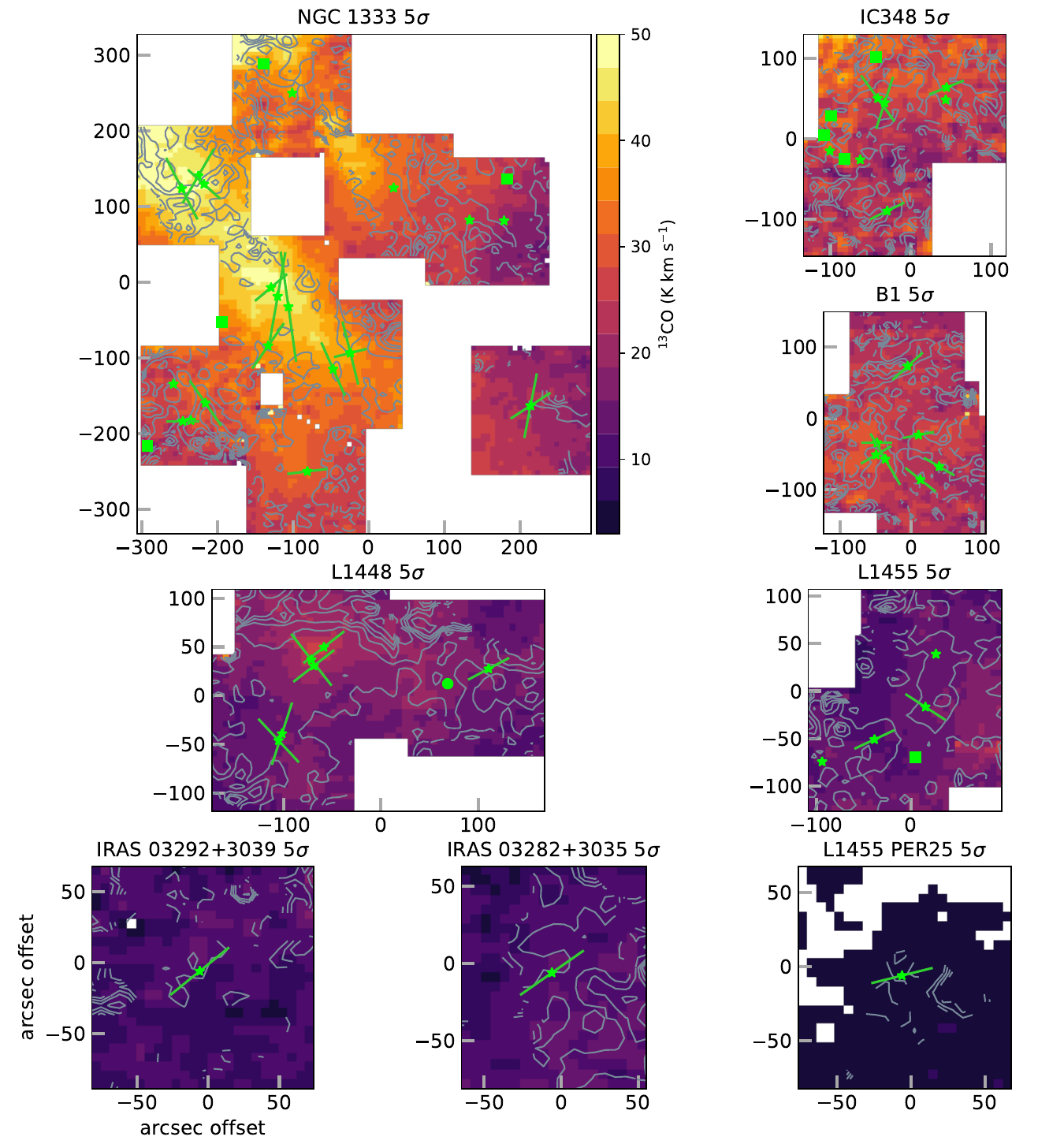}
			\caption{Integrated intensity maps of carbon monoxide \ce{^{13}CO} 1--0 toward five subregions in Perseus: NGC1333, L1448, L1455, B1, and IC348. All maps are shown with the same colorscale range for comparison. Maps are clipped to 5$\sigma$. Gas kinetic temperature map is overlaid in contours in steps of 10., 15., 20., 25., 30., 40. K. Star symbols mark the locations of protostellar systems (Table~\ref{tab:source}). Square symbols mark the location of starless cores (Table~\ref{tab:starless}). The circle symbol marks the position of L1448 IRS2E, whose nature is debated. Straight lines indicate outflow directions for the protostellar systems included in the MASSES survey \citep{stephens2017}.}
			\label{fig:13COmom0}
		\end{figure*}

		Intensity integrated maps were produced using the bettermoments python library \citep{teague2018,teague2019}.
		Before generating the intensity integrated maps, the image edges (xy) were cut off in order to reduce noisy pixels which did not contribute significant data.
		The noise level of the maps was calculated using the bettermoments routine \texttt{estimate\_rms} and using 40 channels on either end of the data cube for \ce{HCN}, \ce{N2H+} and \ce{^{13}CO}, and 100 channels for \ce{HCO+}, \ce{HNC} and \ce{C^{18}O}.
		The choice of channels was based on the presence of hyperfine components and line width.
		Due to the complexity of the NGC1333 map, \texttt{estimate\_rms} did not provide a reasonable estimate for the combined maps of \ce{HCO+} and \ce{HNC}.
		Thus the noise level was estimated from the individual maps instead of the combined map, resulting in noise levels of 0.98 and 0.82 K for \ce{HCO+} and \ce{HNC}, respectively.
		The noise level for the other regions using the above method was in good agreement with the \texttt{estimate\_rms} value.
		A channel mask was then applied to the data cube to select channels with relevant line emission.
		The channel range for \ce{N2H+} and \ce{HCN} includes all hyperfine components.
		Values above 40 K and below -5 K were clipped using \texttt{get\_threshold\_mask} to avoid any pixels that would generate artifacts. 
		The upper limit was based from the \ce{CO} isotopologue maps, which present the brightest emission.
		The intensity integrated maps were then generated with \texttt{collapse\_zeroth}.
		Finally, the maps were clipped to 3$\sigma$ or 5$\sigma$ depending on the integrated intensity peak and are termed the clipped integrated intensity maps in this work (e.g., Section~\ref{subsec:gasT}). 
		The $\sigma$ value was taken from the uncertainty map generated by bettermoments.
		The resulting maps are shown in Figures~\ref{fig:HCNmom0} to \ref{fig:13COmom0}.
		Here a brief description of each map is given.
		
		Hydrogen cyanide (\ce{HCN} 1--0, Fig.~\ref{fig:HCNmom0}) is present in all surveyed regions and shows a clumpy distribution. 
		The \ce{HCN} emission is brightest in NGC1333 and L1448 with peaks toward NGC1333 IRAS4A (28 K~km~s$^{-1}$) and L1448 N (19 K~km~s$^{-1}$). 
		Bright \ce{HCN} peaks ($\sim$30 K~km~s$^{-1}$) are observed west of NGC1333 IRAS2A and south of NGC1333 IRAS4A, most likely arising from their respective outflow cavities. 
		The regions IC348, B1, L1455 and IRAS03282 have weaker emission than NGC1333, with peak integrated intensities of 12, 14, 7, 7 K~km~s$^{-1}$, respectively. 
		The isolated systems, Per25 and IRAS03292 show the weakest integrated intensities peaking at $<$2 and 3 K~km~s$^{-1}$, respectively.
		
		Formyl cation (\ce{HCO+} 1--0, Fig.~\ref{fig:HCOmom0}) is detected in all regions. 
		Its emission is brightest in NGC1333, with the strongest emission located around the SVS13 system (peak: 17 K~km~s$^{-1}$) and IRAS6 (peak: 22 K~km~s$^{-1}$). The \ce{HCO+} emission is weaker toward L1448 (peak: 10 K~km~s$^{-1}$), L1455 (peak: 11 K~km~s$^{-1}$) and IC348 (peak: 9 K~km~s$^{-1}$). In L1448 and L1455 the integrated intensity is similar toward all sources, while in IC348 the emission peaks around Per 32 (but not ED366) and the starless cores in the west. Surprisingly, B1 is not bright in \ce{HCO+} with integrated intensities $<$5 K~km~s$^{-1}$, despite presence of young protostars (B1-bN and B1-bS) and active star formation.
		IRAS03282 presents \ce{HCO+} emission peaking at 3 K~km~s$^{-1}$ and extending north to south. The other two isolated systems, Per25 and IRAS03292, present weak and compact \ce{HCO+} emission with intensities $<$2 K~km~s$^{-1}$.
		
		Hydrogen isocyanide (\ce{HNC} 1--0, Fig.~\ref{fig:HNCmom0}) is detected in all regions and toward all sources. 
		The spatial distribution of both isomers, \ce{HNC} and \ce{HCN}, is similar, but \ce{HNC} shows a less clumpy and more smooth morphology than \ce{HCN}.
		NGC1333 and L1448 are equally bright, with the peak integrated intensity of 15 K~km~s$^{-1}$ located around both NGC1333 SVS13 and L1448 N. 
		IC348, B1, L1455 and IRAS03282 have similar integrated intensities (8, 8, 7, 6 K~km~s$^{-1}$ respectively).
		IRAS03292 presents emission peaking at 3 K~km~s$^{-1}$, with the bulk of the emission located to the southwest of the close binary.
		Per25 has weak \ce{HNC} emission with integrated intensity $<$2 K~km~s$^{-1}$.

		Diazenylium (\ce{N2H+} 1--0, Fig.~\ref{fig:N2Hmom0}) is present in all observed regions.
		Similar to \ce{HCN}, \ce{HCO+} and \ce{HNC}, the brightest \ce{N2H+} emission is detected toward NGC1333 and L1448. In NGC1333 the emission is located between SVS13 and RNO 15 FIR peaking at 30 K~km~s$^{-1}$. In L1448, the \ce{N2H+} peak is located near L1448 N B and peaks at 25 K~km~s$^{-1}$.
		In B1, the peak emission (below 15 K~km~s$^{-1}$) is located on the system B1-b, with the rest of the region showing emission around 10 K~km~s$^{-1}$.
		IRAS03292 and IRAS03282 show integrated intensities similar to B1, peaking at 10 K~km~s$^{-1}$. 
		IC348 and L1455 presents \ce{N2H+} emission peaking at 8 K~km~s$^{-1}$, while Per25 has much weaker emission peaking on-source at 5 K~km~s$^{-1}$.
		
		Two carbon monoxide isotopologues, \ce{C^{18}O} 1--0 (Fig.~\ref{fig:C18Omom0}) and \ce{^{13}CO} 1--0 (Fig.~\ref{fig:13COmom0}), were observed and detected in all regions.
		The brightest \ce{C^{18}O} emission is found toward B1, peaking at 9 K~km~s$^{-1}$.
		Toward NGC1333 \ce{C^{18}O} peaks at 8 K~km~s$^{-1}$ and the bulk of the emission is located in an elongated structure stretching from IRAS7, SVS13 and IRAS2. An area of bright emission is detected around the starless core north of IRAS6.
		In L1448, the peak of \ce{C^{18}O} is located near L1448 N-C with an integrated intensity peak of 6 K~km~s$^{-1}$.
		No peak of \ce{C^{18}O} is detected toward any of the sources in IC348. The maximum integrated intensity is 6 K~km~s$^{-1}$.
		Similarly, L1455, IRAS03282 and IRAS03292 do not show on-source \ce{C^{18}O} peaks, with maximum integrated intensity of 3 K~km~s$^{-1}$ in the diffuse emission around the protostars.
		The \ce{C^{18}O} emission toward Per25 is located on-source and extended around the source, peaking at 1 K~km~s$^{-1}$. 
		\ce{^{13}CO} emission is brightest in NGC1333 along the elongated structure observed in \ce{C^{18}O}, peaking at 49 K~km~$^{-1}$ northeast of IRAS7. A similar peak is detected around the starless core north of IRAS6, consistent with the \ce{C^{18}O}.
		IC348 and B1 present bright emission peaking at 30 K~km~s$^{-1}$ on and around the sources, while L1455 peaks 25 K~km~s$^{-1}$ away from the sources. 
		In L1448, the bulk of the \ce{C^{18}O} is located around L1448 N, peaking at 26 K~km~s$^{-1}$.
		Both IRAS03292 and IRAS03282 peak at 10 K~km~s$^{-1}$, while Per25 peaks at 5 K~km~s$^{-1}$.
		
		Both \ce{CO} isotopologues show different structures to those traced by the other molecular species (\ce{HCN}, \ce{HNC}, \ce{HCO+} or \ce{N2H+}), and do not show a particular distribution relative to protostellar multiplicity (Fig.~\ref{fig:C18Omom0} and \ref{fig:13COmom0}).
		This demonstrates that \ce{CO} is not the best tracer to examine the physical factors that influence star formation and evolution.
		The strongest \ce{^{13}CO} emission is found in NGC1333 extending northeast to southwest, passing along the IRAS7, SVS13 and IRAS2 cores. 
		On the other hand, the brightest \ce{C^{18}O} emission is observed toward the main part of B1, but not IRAS03292 and IRAS03282.
		The observed effects could be due to the optical depth of the \ce{^{13}CO} emission.
		The strong \ce{C^{18}O} emission in B1 could be caused by higher masses in the region.
		However, we note that the \ce{N2H+} and \ce{HCO+} gas masses in B1 are above the average, but are not the highest for all the sources surveyed in the Perseus molecular cloud.
		Hence the strong \ce{C^{18}O} emission in B1 could be product of a chemical effect as well.

	\begin{figure*}
		\centering
		\includegraphics[width=0.95\linewidth]{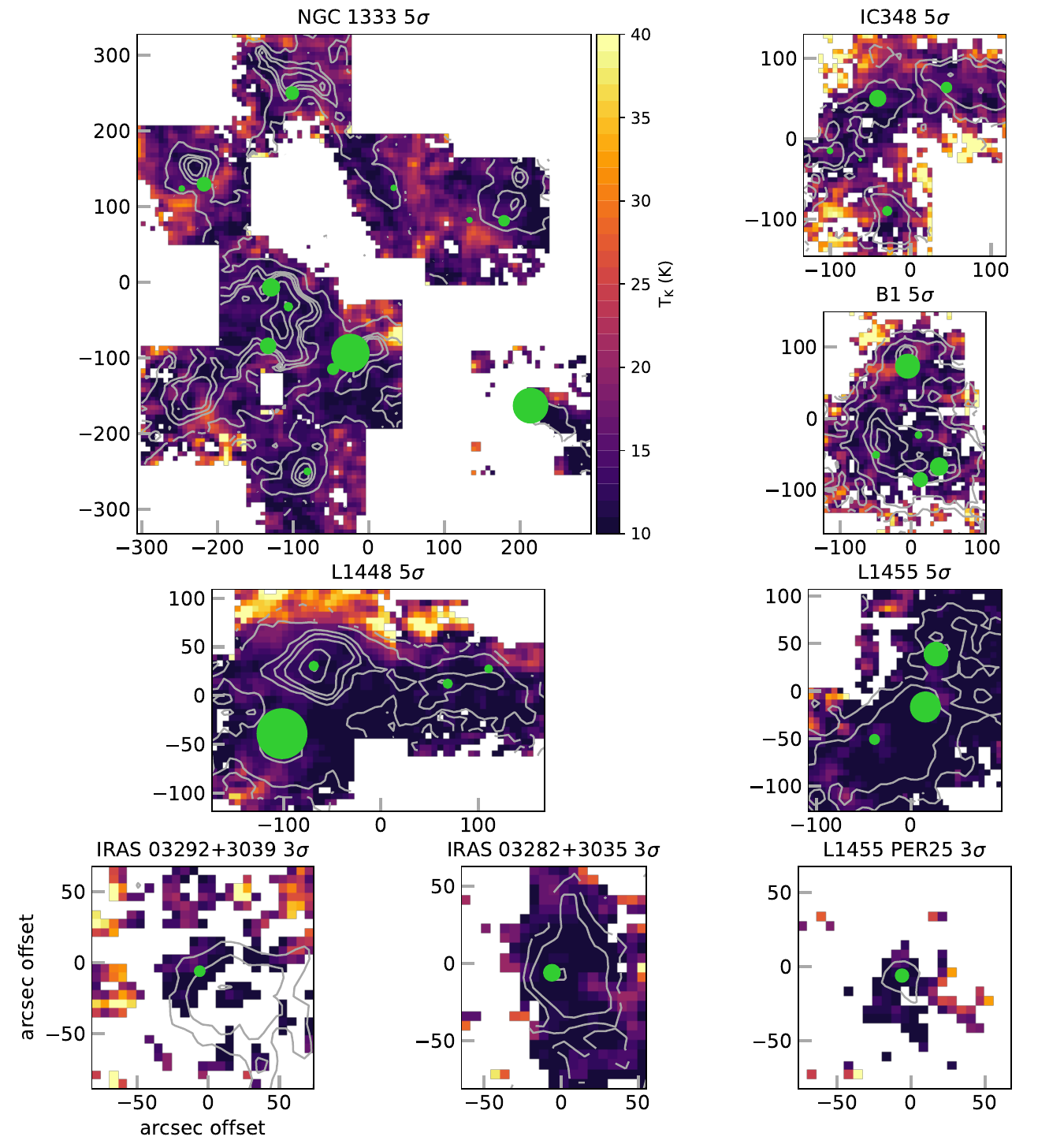}
		\caption{Intensity integrated maps of diazenylium \ce{N2H+} 1--0 (purple contours, see also Fig.~\ref{fig:HCOmom0}) overlaid on the kinetic temperature map (colorscale, same as Fig.~\ref{fig:TKmap}).  Contours are in steps of 1, 2, 4, 6, 12, 14 and 16 K~km~s$^{-1}$. All maps are shown with the same colorscale range for comparison. Black filled circles are proportional to the derived \ce{H2} density for each protostellar system. Derived densities were normalized to the largest value, found toward L1448 C. The normalized value was multiplied by an arbitrary value to make the filled circles visible on the map. The panel size does not influence the size of the filled circles.}
		\label{fig:TKmap_H2density}
	\end{figure*}
	
	\section{Additional statistics plots}
	\label{ap:stats}
\begin{figure*}
		\centering
		\includegraphics[width=0.95\linewidth]{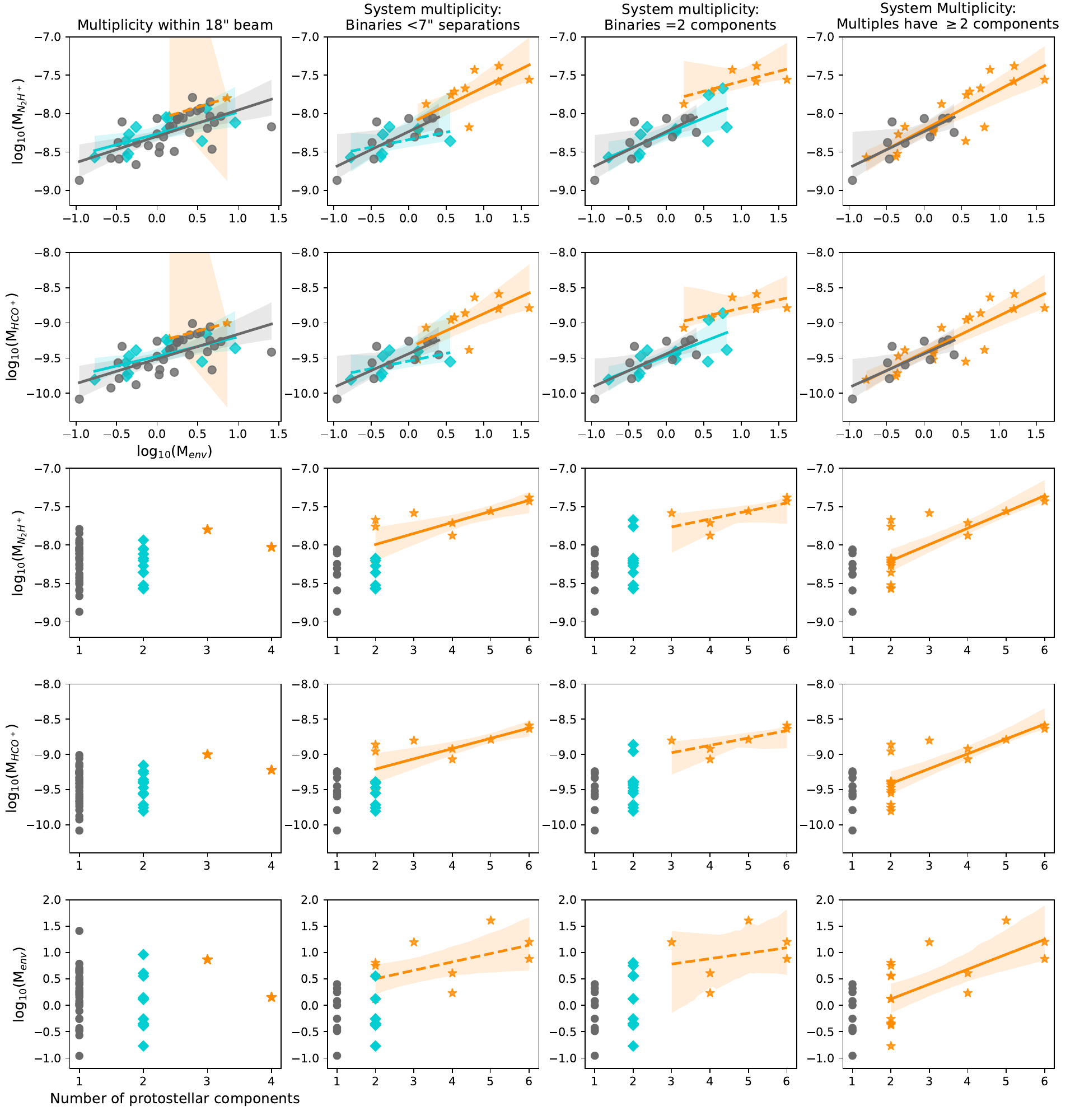}
		\caption{ Relations between envelope mass M$_{env}$ and \ce{N2H+} gas mass (M$_{gas}$(\ce{N2H+}), \textit{first row}, same as Fig.~\ref{fig:massColDen}), and \ce{HCO+} gas mass (M$_{gas}$(\ce{HCO+}), \textit{second row}). Bottom three rows show number of components versus gas mass (M$_{gas}$(\ce{N2H+}), \textit{third row}; M$_{gas}$(\ce{HCO+}), \textit{fourth row}), and versus envelope mass (M$_{env}$, \textit{fifth row}).
			The orange stars show multiple systems, the cyan diamonds show binary systems, and the gray circles show single protostellar systems. Each panel represents one of four ways of grouping the sample presented in this work, and their corresponding correlations. The lines and shaded areas show the linear regression for the data with the corresponding color. Solid lines indicate statistically significant correlations (Pearson $r$ and Spearman $\rho$ p-values$<$0.05), while dashed lines show subsamples with p-values$>$0.05. See Section~\ref{subsec:stats} for discussion on the figure.} 
		\label{fig:numcomp}
	\end{figure*}

	Plot of all gas and dust masses that show significant correlations in the four classification schemes described in Section~\ref{subsec:stats} are shown in Figure~\ref{fig:numcomp}. In addition, the correlations between masses and number of components are also shown. Classification scheme 4 (third column in Fig.~\ref{fig:numcomp}) visually shows a correlation between mass and number of components, albeit flatter than in the other classification schemes. However, the Pearson $r$ and Spearman $\rho$ do not reflect a statistically significant correlation.
	
	\end{appendix}
	
	\bibliographystyle{aa}
	\bibliography{MolCloud_TempDensity_paper1.bib}
	
\end{document}